\documentclass[12pt]{article}
\usepackage{amsmath}
\usepackage{amssymb}
\usepackage{latexsym}
\usepackage{epsfig}
\usepackage{graphicx}
\usepackage{slashed}

\usepackage{amssymb}
\usepackage{euscript}

\catcode`\@=11 \@addtoreset{equation}{section}

\newcommand\be{\begin{equation}}
\newcommand\ee{\end{equation}}
\newcommand{\raw}{\rightarrow}

\newcommand\mathC{\mkern1mu\raise2.2pt\hbox{$\scriptscriptstyle|$}
        {\mkern-7mu\rm C}}              
\newcommand{\mathR}{{\rm I\! R}}         

\begin{document}

\begin{center}
{\large\bf Renormalization for Philosophers}
\end{center}

\begin{center}
J. Butterfield and N. Bouatta \\
Saturday 14 June 2014 \\
\end{center}

\begin{center}
Forthcoming (slightly abridged) in {\em Metaphysics in Contemporary Physics}: a volume of {\em Poznan Studies in Philosophy of Science}, eds. T. Bigaj and C. W\"{u}thrich
\end{center}

\begin{abstract}
We have two aims. The main one is to expound the idea of renormalization in quantum field 
theory, with no technical prerequisites (Sections \ref{tradnut} and \ref{modnut}). Our motivation is that renormalization is undoubtedly one of the great ideas---and great 
successes---of twentieth-century physics. Also it has strongly influenced, in diverse ways, 
how physicists conceive of physical theories. So it is of considerable philosophical 
interest. Second, we will briefly relate renormalization to Ernest Nagel's account of 
inter-theoretic relations, especially reduction (Section \ref{Nag}).

One theme will be a contrast between two approaches to renormalization. The old approach, 
which prevailed from ca 1945 to 1970, treated renormalizability as a necessary condition for 
being an acceptable quantum field theory. On this approach, it is a piece of great good 
fortune that high energy physicists can formulate renormalizable quantum field theories that 
are so empirically successful. But the new approach to renormalization (from 1970 onwards) 
explains why the phenomena we see, at the energies we can access in our particle 
accelerators, are described by a renormalizable quantum field theory. For whatever 
non-renormalizable interactions may occur at yet higher energies, they are insignificant at 
accessible energies. Thus the new approach explains why our best fundamental theories have a 
feature, viz. renormalizability, which the old approach treated as a selection principle for 
theories.

That is worth saying since philosophers tend to think of scientific explanation as only 
explaining an individual event, or perhaps a single law, or at most deducing one theory as a 
special case of another. Here we see a framework in 
which there is a space of theories. And this framework is powerful enough to deduce that 
what seemed ``manna from heaven'' (that some renormalizable theories are empirically 
successful) is to be expected: the good fortune is generic.

We also maintain that universality, a concept stressed in renormalization theory, is 
essentially the familiar philosophical idea of multiple realizability; and that it causes no 
problems for reductions of a Nagelian kind.

 \end{abstract}

 \newpage
\tableofcontents
\newpage

\section{Introduction}\label{intro}
We have two aims. The main one is to expound the idea of renormalization in quantum field 
theory, with no technical prerequisites (Sections \ref{tradnut} and \ref{modnut}). Our 
motivation is that renormalization is undoubtedly one of the great ideas---and great 
successes---of twentieth-century physics. Also it has strongly influenced, in diverse ways, 
how physicists conceive of physical theories. So it is of considerable philosophical 
interest. Second, we will briefly relate renormalization to Ernest Nagel's account of 
inter-theoretic relations, especially reduction (Section \ref{Nag}).\footnote{That Section 
is brief because one of us (JB) discusses Nagelian themes more fully in a companion paper (2014). From now on, it will be clearest to use `we' for a contextually indicated community e.g. of physicists, as in `our best  physical theories'; and `I' for the authors, e.g, in announcing a plan for the paper like `In Section 2, I willÉ'.}

One main point will turn on a contrast between two approaches to renormalization. The 
traditional approach was pioneered  by Dyson, Feynman, Schwinger and Tomonaga in 1947-50: 
they showed how it tamed  the infinities occurring in quantum electrodynamics, and also 
agreed with experiments measuring effects  due to vacuum fluctuations in the electromagnetic 
field---even to several significant figures. After these triumphs of quantum 
electrodynamics, this approach continued to prevail for two decades. For this paper, the main point 
is that it treats renormalizability as a necessary condition for being an acceptable quantum 
field theory. So according to this approach, it is a piece of great good fortune that high 
energy physicists can formulate renormalizable quantum field theories that are so 
empirically successful; as they in fact did, after about 1965, for forces other than 
electromagnetism---the weak and strong forces.

But between 1965 and 1975, another approach to renormalization was established by the work 
of Wilson, Kadanoff, Fisher etc. (taking inspiration from ideas in statistical mechanics as 
much as in  quantum field theory). This approach explains why the phenomena we see, at the 
energies we can access in our particle accelerators, are described by a renormalizable 
quantum field theory. In short, the explanation is: whatever non-renormalizable interactions 
may occur at yet higher energies, they are insignificant at accessible energies.  Thus the 
modern approach explains why our best fundamental theories have a feature, viz. 
renormalizability, which the traditional approach treated as a selection principle for 
theories. (So to continue the metaphor above: one might say that these theories' infinities 
are not just tamed, but domesticated.)

That point is worth making since philosophers tend to think of scientific explanation as 
only explaining an individual event, or perhaps a single law, or at most deducing one theory 
as a special case of, or a good approximation of, another. This last is of course the core 
idea of Nagel's account of inter-theoretic reduction. The modern approach to renormalization 
is more ambitious: it explains, indeed deduces, a striking feature (viz. renormalizability) 
of a whole class of theories. It does this by making precise mathematical sense of the ideas 
of a {\em space} of theories, and a flow on the space. It is by analyzing this flow that one 
deduces that what seemed ``manna from heaven'' (that some renormalizable theories are 
empirically successful) is to be expected: the good fortune we have had is generic. 

But I will urge that this point is not a problem for Nagel's account of inter-theoretic 
relations. On the contrary: it is a striking example of the power of Nagelian reduction. And 
I will end with an ancillary point, which also strikes a Nagelian note. I will argue that 
universality, a concept stressed in renormalization theory, is essentially the familiar 
philosophical idea of multiple realizability; and I will claim (following Sober, Shapiro and 
others) that multiple realizability does not cause problems for reductions of a Nagelian 
kind. 

The plan is as follows. I sketch the old and new approaches to renormalization in Sections 
\ref{tradnut} and \ref{modnut}.\footnote{I will give very few references to the technical 
literature; as perhaps befits a primer. But I  recommend: (i) Baez's colloquial  
introductions  (2006, 2009), of which Sections \ref{tradnut} and \ref{modnut} are an 
expansion into more academic prose; (ii) Wilson's {\em Scientific American} article (1979) 
and Aitchison (1985)'s introduction to quantum field theory, especially its vacuum, which 
discusses renormalization in Sections 3.1, 3.4, 3.6, 5.3, 6.1; (iii) Teller (1989)  and 
Hartmann (2001) as philosophical introductions; (iv) Kadanoff's masterly surveys (2009, 
2013), which emphasize both history and aspects concerning condensed matter---here treated 
in Section \ref{smperspec}.} Then in Section \ref{Nag}, I shall maintain that these 
developments accord with Nagel's doctrines.

\section{Renormalization: the traditional approach}\label{tradnut}
\subsection{Prospectus: corrections needed}\label{tradnutprosp}
Consider a classical point-particle acting as the source of a gravitational or electrostatic 
potential. There is no problem about using the measured force $F$ felt by a test-particle
at a given distance $r$ from the source, to calculate the mass or charge (respectively) of 
the source particle.

Thus in the electrostatic case, for a test-particle of unit charge, the force is given by 
minus the derivative of the potential energy $V$ with respect to the distance $r$ between 
the source and the test-particle. In symbols, this is, for a source of charge $e$ 
(neglecting constants): $F = - \nabla V(r) \sim - \nabla - e/r \equiv -e/r^2$ . We then 
invert this equation to calculate that the source's charge is: $e  = - F.r^2$. (Adding in 
the constants: $F = \frac{-e}{4 \pi \varepsilon_0 r^2}$, where $\varepsilon_0$ is the 
permittivity of free space (electric constant), implies that $e  = - F (4 \pi \varepsilon_0 
r^2)$.)

This straightforward calculation of the source's mass or charge does not work in quantum 
field
theory! There are complicated corrections we must deal with: perhaps unsurprisingly, since 
it amounts to trying to characterize one aspect of an interacting many-field system in a way 
that is comparatively simple and independent of the rest of the system. The corrections will 
depend on the energy and-or momentum with which the test particle approaches the source. A 
bit more exactly, since of course the test particle and source are equally minuscule: the 
corrections depend on the energy (or momentum) with which we theoretically describe, or 
experimentally probe, the system I called the `source'. We will write $\mu$ for this energy; 
and the corrections depending on $\mu$ will be centre-stage in both the traditional and the 
modern approaches to renormalization (this Section and the next).

This Section lays  out the traditional approach in four subsections. In Section \ref{rencc}, 
I introduce the idea that the medium or field around the source can affect the observed 
value of its mass or charge. In quantum field theory, this is often expressed in the jargon 
of ``virtual states'' or `` virtual particles''. Again: it is a matter of the energy-scale 
$\mu$ at which we describe or probe the source.\\
\indent Then in Section \ref{cut}, I report that to get finite predictions, quantum field 
theory needs a regularization scheme. The archetypal scheme is to neglect energies above a 
certain value $\Lambda$; equivalently, one neglects variations in fields that occur on a 
spatial scale smaller than some small length $d$. I also adopt length as the fundamental 
dimension, so that I express regularization as a cut-off length $d$, rather than an energy 
$\Lambda$.\\
\indent In Section \ref{dto0}, I present the core task of the traditional approach to 
renormalization. Since the theory assumes spacetime is a continuum, while $d$ is our 
arbitrary choice,  we need to show consistency while letting $d$ tend to 0. That is: we must 
find an assignment of intrinsic charges (electric charge, mass etc.: called {\em bare 
coupling constants}), to the sources, as a function of the diminishing  $d$, which delivers 
back the  observed value of the charges: i.e. the values we in fact measure at the 
energy-scale $\mu$ at which we probe the system. These measured values are called the {\em 
physical coupling constants}. If we can do this, we say our theory is {\em 
renormalizable}.\\
\indent This requirement is weak, or liberal, in two ways. First: we even allow that the 
assigned intrinsic charge is infinite at the limit $d \raw 0$. (It is this allowance that a 
bare coupling constants be infinite that makes many---including great physicists like 
Dirac---uneasy.)\\
\indent Second (Section \ref{nonren}): we allow that we might have to add other terms to our 
theory (to be precise: to the Lagrangian or Hamiltonian), in order to make a consistent 
assignment. But we only allow a finite number of such terms: this reflects the fact that our 
framework of calculation is perturbative.\\
\indent Then in Section \ref{dyson}, I report (the simplest rendition of) Dyson's criterion 
for when a theory is renormalizable: the dimension (as a power of length) of the bare 
coupling constant(s) needs to be less than or equal to zero. Finally, I report the happy 
news that our theories of the electromagnetic, weak and strong forces are in this sense 
renormalizable. {\em Why} we should be so fortunate is a good  question: which, as announced 
in Section \ref{intro}, I will take up in Section \ref{modnut}.

\subsection{Renormalizing a coupling constant}\label{rencc}
Underlying the details to come, there is a simple idea, which is valid classically and 
indeed an everyday experience. Imagine a ping pong ball under the surface of the water in a 
bath. It is buoyant: in terms of the gravitational field, it has a negative mass. So the 
idea is:
the medium in which a system is immersed can alter the parameters associated with the 
system, even the parameters like an electric or gravitational charge, i.e. its coupling 
constants.

Agreed, in this example we of course retain the notion that the ball has a positive 
intrinsic mass, not least because it can be taken out of the water and then will fall under 
gravity. But three factors make this notion more problematic in fundamental physics, 
especially in quantum theory.\\

\indent (i): We cannot take the system out of the field of force, which is all-pervasive, 
though varying in strength from place to place.\\
\indent  (ii): Even in classical physics, there are important differences between the 
electromagnetic and gravitational fields: differences that make it  wrong, or at least more 
questionable, to conceptually detach an electric charge from the electromagnetic field, than 
it is to detach a mass from the gravitational field. In short, the difference is that the 
gravitational field, but not the electromagnetic field, can be thought of as a mere 
mathematical device giving the disposition of a test-particle to accelerate: cf. the field 
$V$ above. On the other hand, the electromagnetic field has energy and momentum, and effects 
propagate through it at a finite speed: this leads to subtle issues about the self-energy of 
a classical charged particle.\footnote{For these issues, cf. Zuchowski (2013) and references 
therein. Broader philosophical aspects of classical fields are discussed by Hesse (1965), 
McMullin (2002) and Lange (2002).} \\
\indent (iii): In quantum field theory, matter is represented by a quantum field, just as 
radiation (electromagnetism) is. The matter is represented by a fermionic field; e.g. in 
quantum electrodynamics, the electron field. And interactions (forces) between matter are 
represented by a bosonic field; e.g. in quantum electrodynamics, the quantized 
electromagnetic field, whose excitations are photons. In short: the physical system is an 
interacting many-field system, so that it makes little sense to conceptually detach one of 
the fields from the others. All the more so, if we think of our fields as effective, not 
fundamental: I return to this in Section \ref{modnut}.\\

In short: we need to take seriously, in our theoretical description as much as our 
experimental practice, that the system of interest, e.g. an electron (or excitation of the 
electron field), is immersed in a wider system through which we ``access'' it. This has two 
aspects which we need to spell out. The second is  more briefly stated, and is fundamental: 
it will dominate the sequel. But the first sets the stage.

\subsubsection{Virtual particles and perturbation theory}\label{vpptbn}
First, we need the ideas of: {\em virtual states}, also often called {\em virtual 
particles}, which arise in the context of {\em perturbation theory}.

 In quantum theory, we typically solve a problem by finding the states of definite energy 
and their corresponding values of energy. These are the energy eigenstates, i.e. eigenstates 
of the Hamiltonian (energy-function), and their eigenvalues. For once these are known, most 
of what we might want to know can be calculated. Usually, we cannot exactly calculate the 
eigenstates: the Hamiltonian is intractable. (The Hamiltonian $H$ is essentially a matrix, 
and calculating the eigenstates and eigenvalues is a matter of changing  bases in the vector 
space of states so as to render the Hamiltonian matrix diagonal i.e. to make all 
non-diagonal entries zero.) But often enough, $H$ is ``close'' to another Hamiltonian, $H_0$ 
say, which {\em is} tractable, in that we can calculate $H_0$'s eigenstates. Here, closeness 
means, roughly, that there is an additional Hamiltonian matrix $H_i$ such that  $H = H_0 + 
\varepsilon H_i$ where $\varepsilon$ is a small number. Since $\varepsilon$ is  small,  
$H_0$ and $H$ are approximately equal, $H_0 \approx H$.  We can then write the desired
eigenstates of $H$ as superpositions (weighted sums of) of the eigenstates of $H_0$ (which, 
recall, we can calculate). Thus $| \psi_a \rangle = \Sigma_j c_j | \psi^0_j \rangle$: where 
$a$ labels the real Hamiltonian's eigenvalue (meaning just that $H | \psi_a \rangle = a | 
\psi_a \rangle$); $j$ labels the various eigenvalues of $H_0$, whose eigenstates are the $| 
\psi^0_j \rangle$; and our task is to calculate the complex numbers $c_j$. It is these 
eigenstates of $H_0$ that are called {\em virtual states} or {\em virtual particles}.

This jargon is especially used in quantum field theory, where the real Hamiltonian is 
usually complicated enough to force us to appeal to {\em perturbation theory}. This is a 
general framework for solving intractable problems by treating them as small adjustments to 
tractable problems: e.g. by adding a term, $\varepsilon V$ say, where $\varepsilon$ is a 
small number and $V$ a potential function, to the governing equations of the tractable 
problem. One then tries to calculate the quantity of interest (in our example, one of the 
$c_j$) by expressing it as a power series $\Sigma^{\infty}_n \alpha^n A_n$, where $\alpha$ 
is small, i.e. less than one, so that $\alpha^n \raw 0$ as $n \raw \infty$. Here, $\alpha$ 
may be the original $\varepsilon$, or some related number. The hope is that $\alpha^n$ 
tending to $0$ will make the terms $\alpha^n A_n$ for higher values of $n$ go to 0. If we 
are  lucky, the first few terms of the series will give us an answer that is accurate enough 
for our purposes; and if we are very lucky, the series may even converge to the exact answer 
(i.e. the limit of the successive partial sums $\Sigma^N_n \alpha^n A_n$ is finite and is 
the exact answer). Whether these hopes are realized will of course depend on the $A_n$ not 
growing too quickly.

I should stress immediately that in quantum field theory, the success of this sort of 
perturbative analysis is mixed. On the one hand, there is astounding success: in some cases, 
in our best theories, the first few terms of such a series give an answer that is 
astonishingly accurate. It matches the results of delicate experiments to as much as ten 
significant figures, i.e. one part in $10^{10}$. That is like correctly predicting the 
result of measuring the diameter of the USA, to within the width of a human hair! For 
example, this accuracy is achieved by the prediction in quantum electrodynamics of the 
magnetic moment of the electron; (Feynman 1985, pp. 6-7, 115-119; Schweber 1994, p. 206f.; 
Lautrup and Zinkernagel 1999).

On the other hand, there are serious mathematical problems. It is not just that, in general, 
the power series expressions we use, even in our best theories, are not known to converge, 
and are sometimes known not to converge. There are two deeper problems, which I have not yet 
hinted at.

The first concerns the mathematical definition of the interacting quantum field theory, 
which our perturbative approach with its various power series is aiming to approximate. 
Unfortunately, we do not at present have a rigorously defined interacting quantum field 
theory, for a four-dimensional spacetime. There are such theories for lower spacetime 
dimensions; and there has been much effort, and much progress, towards the goal. But so far, 
it remains unattained. One brief way to put the chief difficulty is to say that the central 
theoretical notion of a quantum field theory, the {\em path integral} (also known as: {\em 
functional integral})---which is what our power series aim to approximate---has at present 
no rigorous mathematical definition, except in special or simple cases such as there being 
no interactions.\footnote{\label{rigorref}{For a glimpse of these issues, cf. e.g. Jaffe 
(1999, 2008), Wightman (1999).}}

The second problem is related to the first; it indeed, it is part of it. But the second 
problem specifically concerns the perturbative approach with its power series, and will be 
centre-stage in this paper. So it is best stated separately. In short: not only do the power 
series usually fail to converge; also, the factors $A_n$ (in the successive terms $\alpha^n 
A_n$) are often {\em infinite}. Thus the worry that the $A_n$ might `grow too quickly' for 
the power series to converge, as I put it above, was a dire under-statement. Nevermind $A_n$ 
being so large for large $n$ that the series might diverge: the problem is really that each 
{\em term} $\alpha^n A_n$ is infinite! This is quantum field theory's notorious {\em problem 
of  infinities}: which, as we will see, is addressed by renormalization. Why the $A_n$ are 
infinite, and how renormalization addresses this by introducing a cut-off and then analysing 
what happens when the cut-off tends to a limit, will be taken up in Section \ref{cut} et 
seq. For the moment, I just confess at the outset that the overall problem of infinities 
will {\em not} be fully solved by renormalization, even by the modern approach (Section 
\ref{modnut}). The infinities will be tamed, even domesticated: but not completely 
eliminated.\footnote{But that is perhaps unsurprising since, as I said, this second problem 
is part of the first. So if it were fully solved, so might be the first problem also.}

As an example of treating interactions in quantum theory using perturbation theory, let us 
consider an electron immersed in, and so interacting with, an electromagnetic field. Here, 
the electron need not be described by a quantum field; it can be described by elementary 
quantum mechanics; but we consider the electromagnetic field to be quantized. We  take it 
that we can solve the electron considered alone: that is, we can diagonalize its Hamiltonian 
$H_e$ say---this is essentially what Schr\"{o}dinger did in 1925.
And we take it that we can solve the field considered alone: that is, we can diagonalize its 
Hamiltonian $H_f$---this is essentially what Dirac did in 1927. But the  interaction  means 
that the  Hamiltonian of the total system, electron plus field,  is not just the 
(tractable!) sum of $H_e$ and $H_f$, call it $H_0$: $H_0 := H_e + H_f$. In terms of 
eigenstates: the energy eigenstates of the total system are not just products of those of 
the electron and the field; and so the total system's energy eigenvalues are not just sums 
of the individual eigenvalues.

But happily, the interaction is rather weak. We can write the total Hamiltonian as $H = H_0 
+ \varepsilon H_i$, where $H_i$ represents the interaction and $\varepsilon$ being a small 
number represents its being weak; and then embark on a perturbative analysis.  In 
particular, we may expand an energy eigenstate in terms of the eigenstates of $H_0$, which 
are each a product of an electron eigenstate and field eigenstate: which latter are states 
with a definite number of photons (i.e. excitations of the field). So according to the 
jargon above: these photons will be called `virtual photons'.\footnote{So NB: `virtual' does 
{\em not} connote `illusory', nor `merely possible', as it does in the jargon of `virtual 
work' etc. in classical mechanics---for which cf. Butterfield (2004).} And as I stressed, 
the theory that treats both the electron and the electromagnetic field as quantum fields 
which interact with each other, i.e. the theory of quantum electrodynamics, is amazingly 
empirically accurate. Such an accurate theory is surely getting something right about 
nature: despite the issues about renormalization, to which we now turn (and some of which, 
as we shall see later in the paper, are not yet resolved).

\subsubsection{Energy scales}\label{scal}
I said, just before Section \ref{vpptbn}, that this second aspect is  more briefly stated 
than the first, but is fundamental and will dominate the sequel. It amplifies the basic 
point I announced at the start of Section \ref{tradnut}: that while in classical physics, 
there seems no problem about using the measured force felt by a test particle so as to 
calculate the charge or mass  (coupling constant) of the source, this straightforward 
approach fails in quantum theory---we need to add complicated corrections.

Thus the general, or ideal, classical situation is that our theory says a measured quantity, 
e.g. a force $F$ on a test particle, is a function of the charge (coupling constant) $g$ of 
the source: $F = F(g)$; (the function being given by our theory, as in the electrostatic 
formula, $F = - \nabla - e/r$). So the task if to measure $F$ and invert this equation to 
calculate $g$ as a function of $F$: $g = g(F)$. But in quantum field theory, this approach 
breaks down: perhaps unsurprisingly, since it amounts to trying to characterize one aspect 
of an interacting many-field system in a way that is comparatively simple and independent of 
the rest of the system: recall (iii) at the start of this Section.

Broadly speaking, the corrections depend on the energy and-or momentum with which the test 
particle approaches the source. A bit more exactly, recognizing that the test particle and 
source are not conceptually distinguished, since e.g. they might both be electrons: writing 
$\mu$ for the energy (or momentum) with which we theoretically describe, or experimentally 
probe, the system, the corrections depend on $\mu$.

So let us write $g(\mu)$ for the {\em physical coupling constant}, i.e. the coupling 
constant that we  measure: more exactly, the coupling constant that we calculate from what 
we {\em actually} measure, in the manner of $g = g(F)$ above, in the simple electrostatic 
example. Then the notation registers that
$g(\mu)$ is a function of $\mu$. But also: it is a function of the {\em bare coupling 
constant}, $g_0$ say, that appears in the theory's fundamental equations (like $e$) in the 
electrostatic example). So we can write $g(\mu) \equiv g(\mu, g_0)$.

Using the details in Section \ref{vpptbn} about virtual states and perturbation theory, we 
can fill this out a bit. The Hamiltonians of our successful interacting quantum field 
theories, such as quantum electrodynamics, are indeed intractable, because they include 
terms (cf. $H_i$ in Section \ref{vpptbn}) for interactions between the various fields, e.g. 
the electron field and the electromagnetic field. So we often analyse problems using 
perturbation theory, and in particular the eigenstates of the free Hamiltonian. Similarly if 
we formulate our theories using the alternative {\em Lagrangian}, rather than Hamiltonian, 
framework. The Lagrangian function (which is essentially a difference of energy functions) 
is intractable, because it contains interaction terms; and so again, we turn to perturbation 
theory.

Usually, for both frameworks and for most problems, perturbation theory yields, as its 
approximate answer to the problem, a power series in the coupling constant, i.e. $\Sigma_n 
g^n A_n$; or a power series in some closely related number. Note that getting a power series 
in the coupling constant is unsurprising, given Section \ref{vpptbn}'s remark that one often 
gets a power series in the parameter $\varepsilon$, which in the interaction term $H_i = 
\varepsilon V$, looks like a coupling constant. Cf. Section \ref{renccdetail} for some 
details.

Besides, as one would guess for a power series in the coupling constant: increasing the 
exponent $n$ from term to term corresponds to  considering corrections which take account of 
successively more interactions between the fields concerned. So the other factor in each 
term, viz. $A_n$, encodes many ways that there can be $n$ such interactions. In particular 
as we will see in Sections \ref{renccdetail} and \ref{cut}: it allows for these interactions 
occurring in various places in space and time (by $A_n$ being an integral over the various 
places, typically over all of space between two times), and at various energies and momenta 
(by integrating over energies and momenta).

Here we connect with the idea of {\em Feynman diagrams}. We think of the interactions as 
occurring between particles, i.e. between excitations of the fields, and we represent the 
idea that the interaction occurs at a point $x$ in spacetime by drawing lines coming from the 
past and reaching to the point, for the incoming particles, and lines from the point towards 
the future, for the outgoing particles. So the point becomes a vertex of a diagram of lines, 
and the diagram corresponding to the term $g^n A_n$ will in general have $n$ vertices. 
Furthermore, since our perturbative analysis essentially involves calculating eigenstates of 
the full interacting Hamiltonian by expanding them in states of the free Hamiltonian: in the 
detailed calculation of the contribution $g^n A_n$ made by a single diagram, the internal 
lines represent propagations of virtual states/particles.\footnote{For some details, cf. 
e.g. Aitchison (1985: Sections 3.4-3.6, 5.3) and Feynman (1985, especially  pp. 115-122). 
They both discuss the phenomenon of {\em vacuum polarization}, and so {\em screening}: the 
intuitive idea is that  $g(\mu)$ will be greater at high energies because the test particle 
penetrates further past the cloud of virtual particles that surround the source, and so 
``feels'' a higher coupling constant. In Section \ref{beta}, we will see the opposite 
phenomenon, {\em anti-screening} or {\em asymptotic freedom}, where $g(\mu)$ is a decreasing 
function  of energy.}

\subsubsection{Some Details}\label{renccdetail}
By way of filling out Sections \ref{vpptbn} and \ref{scal}, here are some details of 
elementary time-dependent perturbation theory. This will be enough to suggest how in a 
quantum field theory we often work with a power series in the coupling constant $g$; and to 
connect a term of such a series with a Feynman diagram. (NB: nothing in this Subsection is 
needed later, so it can be skipped!)

So suppose again that the full Hamiltonian $H$ is $H_0 + H_i \equiv H_0 +  \varepsilon V$, 
where the parameter $\varepsilon$ is small, and we expand in powers of $\varepsilon$. We 
define the {\em interaction picture} by
\be
| {\bar \psi}(t) \rangle := \exp(i H_0 t /\hbar) | \psi(t) \rangle \;\;\ ; \;\;
{\bar V}(t) = \exp(i H_0 t /\hbar) V \exp(-i H_0 t /\hbar) .
\ee
so that the equation of motion of $|{\bar \psi}(t) \rangle$ is
\be
i \hbar \frac{d}{dt} |{\bar \psi}(t) \rangle =  \varepsilon {\bar V}(t)|{\bar \psi}(t) 
\rangle.
\ee
We assume a solution is a power series in $\varepsilon$:
\be
| {\bar \psi}(t) \rangle = \Sigma^{\infty}_{n=0} \; \varepsilon^n | {\bar \psi}_n(t) \rangle 
;
\ee
with $| {\bar \psi}_n(0) \rangle = 0$ if $n > 0$, since at $t=0$ the state $| {\bar \psi} 
\rangle$ is the initial state $|\psi_0 \rangle$ independent of $\varepsilon$.

Substituting the power series into the equation of motion and equating powers of $n$ gives a 
sequence of equations. The $n$th equation is:
\be
i \hbar \frac{d}{dt} |{\bar \psi}_n(t) \rangle =   {\bar V}(t)|{\bar \psi}_{n-1}(t) \rangle.
\ee
These can be solved successively to give:
\be
|{\bar \psi}_n(t) \rangle = \frac{1}{(i \hbar)^n} \int^t_0 dt_n \int^{t_n}_0 dt_{n-1} \cdot 
\cdot \cdot \int^{t_2}_0 dt_1 \; {\bar V}(t_n) \cdot \cdot \cdot {\bar V}(t_1) \; |\psi_0 
\rangle .
\ee
So
\be
|{\bar \psi}(t) \rangle = \Sigma_{n=0} \; (\frac{\varepsilon}{i \hbar})^n \int^t_0 dt_n 
\int^{t_n}_0 dt_{n-1} \cdot \cdot \cdot \int^{t_2}_0 dt_1 \; {\bar V}(t_n) \cdot \cdot \cdot 
{\bar V}(t_1) \; |\psi_0 \rangle .
\ee

Carrying this over to quantum field theory: in short, the factors ${\bar V}(t_i)$ are 
replaced by a product of field operators $\phi$, at a spacetime point $x$, representing an 
interaction between the fields at that point. So a factor replacing ${\bar V}(t_i)$ looks 
like  ${\hat \phi}_1(x_i){\hat \phi}_2(x_i) ... {\hat \phi}_k(x_i)$ to represent $k$ 
different fields interacting at the spacetime point $x_i$. So increasing the exponent $n$ 
from term to term in a power series in the coupling constant,  $\Sigma_n g^n A_n$, amounts 
to  taking account of successively more interactions between the fields concerned. And in 
each term of the power series, the other factor,  viz. $A_n$, encodes many ways that there 
can be $n$ such interactions. In particular, it allows for these interactions:\\
 \indent (i) occurring in various places in space and time, by $A_n$ being a $n$-fold 
multiple integral over the spacetime points ($x_i, i = 1, ..., n$): typically, an integral 
over all of space between two times; and \\
 \indent (ii) occurring at various energies and momenta, by $A_n$ integrating over various 
possible energies and momenta of the interacting particles, especially the intermediate 
virtual particles.\\
As discussed above, the ideas in (i) and (ii) lead to  {\em Feynman diagrams}.

\subsection{The cut-off introduced}\label{cut}
In the closing remark of Section \ref{renccdetail}, that $A_n$ integrates over various 
possible energies and momenta, lurks the notorious problem of quantum field theory's 
infinities: the second problem of Section \ref{vpptbn} which, as announced there, is 
addressed by renormalization.

Typically, $A_n$ is a (multiple) integral) over energy or momentum $k$, extending from some 
value, say $k_0$ (maybe zero), upto infinity of a function that increases at least as fast 
as a positive power of $k$, say $k^a$. So $A_n$ looks like $\int^{\infty}_{k_0} dk \; k^a$. 
If $a > 0$, this integral is infinite; as $k \raw \infty$, so does $k^a$, for positive $a$, 
making the integral infinite.

 So  to get a finite answer from our formulas, we impose a {\em cut-off}: we replace the 
upper limit in the integral, $\infty$,  by a suitably high energy or momentum, written 
$\Lambda$. (There are other `less crude' ways to secure a finite answer---called {\em 
regularizing} the integrals---but I will only consider cut-offs.)

I have introduced the cut-off as an energy $\Lambda$. But in quantum theory, energy is like 
the reciprocal of distance; in the jargon, `an inverse distance': energy $\sim$ 1/distance. 
(And so distance is like an inverse energy.) This is due to the fundamental equations, due 
to de Broglie, relating `particle' and `wave' properties: viz. that momentum $p$ is 
inversely proportional to wavelength $\lambda$ with proportionality constant $h$. That is: 
$p = h/\lambda$. (NB: $\lambda$ and $\Lambda$ are very different: an unfortunate notational 
coincidence, but a widespread one ...)  Wavelength is the number of units of length per 
complete cycle of the wave. So writing $k$ for the reciprocal, called {\em wave-number}, 
i.e. the number of wave-cycles per unit-length, we have: $p = hk$. So high momenta (and high 
energies) correspond to high wave-number, which means short wavelengths and high 
frequencies.

So the cut-off energy $\Lambda$ corresponds, in terms of distance, to a cut-off at a small 
distance $d$.
That is: imposing the cut-off, i.e. requiring by {\em fiat} that $\int^{\infty}_{\Lambda} dk 
\; ... \equiv 0$  means ignoring contributions to the integrand that vary on a distance 
shorter than $d$.  In other words: to get finite answers, we are declaring that the theory  
claims there are no fields varying on scales less than $d$. At least: any such fields do not 
contribute to the specific process we are calculating. Or at least: the theory  claims this, 
unless and until we consider the limit $d \raw 0$---which we will do in Section \ref{dto0}.

So returning to the notation of Section \ref{scal}:   the  physical coupling constant, 
$g(\mu)$, is a function, not only of the bare coupling constant $g_0$ and of $\mu$ itself of 
course, but also of the cut off $d$. Thus:
\be\label{gmu}
g(\mu) \equiv g(\mu, g_0, d).
\ee
So we can now state our task, introduced at the start of Section \ref{tradnut}, more 
precisely. We are to measure $g(\mu)$  (better: to calculate it from what we really measure, 
like the force $F$ in the simple electrostatics example) and then invert eq. \ref{gmu}, i.e. 
write $g_0 = g_0(g(\mu), d)$,  so as to calculate which value of the bare constant would 
give the observed $g(\mu)$, at the given $d$. {\em This task is the core idea of the 
traditional approach to renormalization.}

It is sometimes convenient, for the sake of uniformity, to express all dimensions in terms 
of length. Section \ref{dyson1} will give more details. But for the moment, just note that 
we can trade in the energy-scale $\mu$ for an inverse length, say $\mu \sim 1/L$ where $L$ 
is a length. NB: $L$ {\em is not the cut-off} $d$! We can think intuitively that $L$ is 
experimental, and $d$ is theoretical. That is: $L$ is our  choice about how microscopically 
to describe or probe---to peer into---the system. On the other hand, $d$ is a (generally 
much smaller) length below which we are taking our theory to say there are no contributions. 
So we re-express the physical coupling constant $g(\mu)$ as a function of $L$: we will use 
the same letter $g$ for this function, so that we write $g(L) \equiv g(\mu)$. Thus eq. 
\ref{gmu} becomes:
\be\label{gL}
g(L) \equiv g(L, g_0, d).
\ee

In the next two sections, I turn to two further aspects of the task just described: of 
inverting eq. \ref{gL} and assigning values to $g_0$ that give the observed $g(\mu)$. These 
aspects concern:\\
 \indent \indent (i) letting $d \raw 0$;\\
\indent \indent (ii) needing to allow for some extra terms in the equations: which will 
return us to the analogy of the ping pong ball, at the start of Section \ref{rencc}.

\subsection{Letting the cut-off $d$ go to zero}\label{dto0}
Broadly speaking, the exact value of the cut-off $d$ is up to us. Agreed: for some of the 
troublesome infinities---some of the infinite terms $A_n$ in perturbative analyses of some 
problems---the physics of the problem will suggests a range of values of $d$ that are 
sensible to take. That is: the physics suggests that no phenomena on scales much smaller 
than $d$ will contribute to the process we are analysing. One such example, very famous for 
its role in the establishment of quantum electrodynamics, is the Lamb shift, where the 
electron's Compton wavelength seems a natural lower limit to $d$; cf. Aitchison (1985, 
Section 3.5-3.6), Schweber (1994, pp. 223-247).

But of course, we would like the theory and its predictions to be independent of any human 
choice. So generally speaking, it is natural to take $d$ smaller and smaller, at fixed 
$\mu$; and  to consider how $g_0$ varies as a function of $d$, while preserving the observed 
$g(\mu)$.

More precisely: if we believe that:\\
\indent \indent (a) spacetime is continuum, and \\
\indent \indent (b) our theory holds good in principle at arbitrarily short lengths,\\
then we surely also believe that at fixed $\mu$ (or at least at some, maybe theoretically 
judicious or appropriate, $\mu$: such as the observed $\mu$), $g_0$ goes to a limit: that 
is:
\be\label{limg0}
{\mbox{ there exists }}\; \; \lim_{d \raw 0; (g({\mu}) \; \mbox{fixed at observed value})} 
g_0 \;.
\ee
We will later treat the issues that (i) since we can vary $\mu$, there are various observed 
values $g(\mu)$, and therefore (ii) whether we should require eq. \ref{limg0} for {\em all} 
the observed values $g(\mu)$. We will also treat letting $\mu$ go beyond the observed range, 
even letting it go to infinity, although we cannot measure $g(\mu)$ above some technological 
(financial!) maximum value of $\mu$.\\

If the limit in eq. \ref{limg0} exists and is finite, i.e. $\in \mathR$, we say: {\em the 
theory is finite}. As the label suggests: in the face of the troublesome infinities, such a 
conclusion would be a relief. But I should stress some limitations of this conclusion. There 
are three obvious general ones; (cf. the references in footnote \ref{rigorref}):\\
\indent (i): this limit being finite does not imply that {\em any} of the power series which 
our perturbative analysis provides for specific physical problems converges;\\
\indent (ii): even if such a series, for a problem at a given value (or range) of $\mu$, 
does converge, this does not settle the behaviour of the corresponding series at other 
$\mu$; and that behaviour might be bad---in particular, arbitrarily high $\mu$ might produce 
a troublesome infinity;\\
\indent (iii): even if all goes well in a perturbative analysis, i.e. the various series 
converge for the ranges of parameters for which we might hope, there remains a gap between a 
perturbative analysis and the original physics problem or problems.\\

But even with these prospective limitations, some successful quantum field theories are {\em 
not} finite. The paradigm case is QED. For QED, the limit in eq. \ref{limg0} is infinite. 
That is: for arbitrarily high cut-offs, the bare charge $g_0$ is  arbitrarily high (and 
remains so for yet higher cut-offs). Mathematically, this is like elementary calculus where 
we say that some function $f(x)$ tends to infinity as $x$ tends to infinity,
e.g. $\lim_{x \raw \infty} \surd x = \infty$. But of course this last is `just' the infinity 
of pure mathematics. But here we face (assuming (a) and (b) above) a {\em physically real 
infinity} viz. as the value of the bare coupling constant.

The consensus, on the traditional approach to renormalization, is that this physically real 
infinity is {\em acceptable}. After all: since by construction we do not measure (nor 
calculate from measurements) the bare constant, we do not need to allow an `Infinity' 
reading on our apparatus' dials. To reflect this consensus, the adjective `renormalizable', 
with its honorific connotations, is used. That is: If the limit in eq. \ref{limg0} exists, 
albeit perhaps being $\pm \infty$, we say the theory is {\em renormalizable}. So in 
particular: QED is renormalizable in this sense, though not finite. (This definition of 
`renormalizable' will be filled out in the next two subsections.)

But I should add that despite this consensus, most physicists would admit to some discomfort 
that the bare constant should be infinite in the continuum theory that, according to (a) and 
(b), we are thinking of as fundamental. Thus great physicists like Dirac have been very 
uncomfortable (cf. the citations in Cao (1997, pp. 203-207)); and Feynman himself calls 
renormalization `a dippy process' and `hocus-pocus' (1985, p. 128);  (Teller (1989) is a 
philosophical discussion).

Besides, this discomfort does not depend on believing exactly (a) and (b) above. Suppose 
that instead we merely believe the corresponding claims of possibility:\\
\indent \indent(a') spacetime might be a continuum, and \\
\indent \indent (b') our theory should be a consistent description (of the interactions in 
question) at arbitrarily short lengths.\\
In short: we merely believe that our theory, formulated with a continuum spacetime as 
background, is a ``way the world could be''. Then we surely are committed to believing that 
at fixed $\mu$ (or at least some, maybe theoretically judicious or appropriate, $\mu$), 
$g_0$ goes to a limit. And again, it is uncomfortable that this limit is infinity. Although 
this yields, not an {\em actual} physical  infinity, but only a {\em possible} physical  
infinity: surely a philosopher should be uncomfortable at such a possibility. (Section \ref{eft?} will return to this, suggesting a way to ease the discomfort.)\\

But despite this discomfort, the fact remains that after facing the troublesome infinities, 
it is obviously a great intellectual relief to find one's theory to be renormalizable, even 
if not finite. It means we can succeed in our task: namely, to consistently assign bare 
constants (albeit perhaps infinite ones) so as to recover the observed physical 
coupling---and do so independently of the cut-off $d$ we adopt so as to make integrals 
finite.

 This relief prompts the idea that even if one does not explicitly endorse (a) and (b) (or 
perhaps, even (a') and (b')), one should adopt renormalizability as a reasonable {\em 
criterion for selecting theories}. Thus the idea is: a  good theory of whatever 
interactions, should make sense, albeit perhaps with an infinite
bare coupling constant, when formulated with a continuum spacetime as background. This is 
indeed how renormalizability was regarded in the traditional approach to renormalization, 
which reigned ca. 1950 to 1970: acceptable theories should be remormalizable.

\subsection{The need for extra terms}\label{nonren}
The main issue in trying to write down a renormalizable theory is that we may need to add 
(to the Lagrangian or Hamiltonian function) one or more  terms  to represent extra 
interaction(s) between the fields, even though we believe the bare coupling constant for the 
extra interaction(s) are zero. I first describe (i) the physical rationale for this; and 
then (ii) the refinement it prompts in the definition of renormalizability, and thereby in 
what the task of formulating a renormalizable theory involves.

\subsubsection{The physical rationale}\label{physrat}
Thus suppose we think the bare coupling constant of some interaction is zero. That is, we 
think that in our fundamental theory, a certain interaction---which would typically be 
represented by some product of field operators---does not happen. Then we will be tempted to 
have no term representing this interaction in our theory (as a summand in our Lagrangian or 
Hamiltonian). For whatever the form, say $\cal F$, of the interaction (i.e. of the product 
of operators), we will think we can leave the term out of all equations. For if $g = 0$ then 
the term $g . {\cal F}$ equals zero, and surely contributes nothing to any equations.

But this might be a mistake! Despite the zero {\em bare} coupling, the interaction might 
have a non-zero {\em physical} coupling constant at some scale $\mu$; i.e. $g(\mu) \neq 0$. 
Indeed, this situation can arise not only for:\\
\indent (a): the strength of a certain interaction between given fields; but also for\\
\indent  (b): the mass or charge of a given field, or as people say: the mass or charge of a 
given particle (treated as an excitation of a field). \\
In  case (b), we would be tempted to omit as pointless terms for all possible interactions 
of the given field (particle) that depend on that mass or charge, since the terms are 
apparently zero, and so surely contribute nothing to any equations. But this might be a 
mistake: the physical coupling constant may be non-zero at some scale $\mu$. In such a case, 
we say: `the field (or particle) acquires a mass/charge at the scale 
$\mu$'.\footnote{Intuitively, case (b) seems more problematic than case (a): for the mass or 
charge of a given field seems more ``intrinsic'' to it than is participation in a certain 
interaction with other fields, and our habituation to mass and charge in classical physics 
makes us think such properties are ``given'' prior to any interactions, rather than acquired 
from them.}

The  analogy of the ping pong ball, mentioned at the start of Section \ref{rencc}, may help. 
There, the fact that it falls in a vacuum (or air) but is buoyant in water---i.e. exhibits a 
positive gravitational mass in vacuum and air, but a negative one in water---illustrated the 
general idea that the coupling constants associated to a system can be altered by the medium 
in which the system is immersed. But now imagine the ping pong ball is so light as to be 
massless in air, so that in air, it does not fall under gravity but floats, weightless; yet 
when immersed in water, it `acquires a mass' (in fact a negative one, since it moves 
upwards, opposite to the force of gravity). Thus a system with $g_0 = 0$ might have at some 
scale $\mu$ a non-zero physical coupling constant, $g(\mu) \neq 0$, which you could measure; 
(better: calculate from actual measurements).

So faced with this situation, whether case (a) or case (b): we should of course include the 
corresponding term or terms in our fundamental equations. For recall that our basic task is 
to find values of the bare constants that (at the given $\mu$ and $d$) imply the measured 
values of $g(\mu)$. Here it will help to generalize the notation slightly to reflect the 
fact that there are of course several, even many, coupling constants to consider; as well as  
several, even many, possible interactions (terms in the Lagrangian or Hamiltonian that are 
typically products of operators). So suppose that
there are in all $N$ physical coupling constants, $g_1(\mu), g_2(\mu), ... g_N(\mu)$, 
occurring in the various terms/interactions in our theory. Then we cannot expect to have 
them implied by less than $N$  bare constants, even if we think some of the bare constants 
are zero. After all, to fit $N$ numbers, we expect to need $N$ numbers.

\subsubsection{A refined definition of renormalizability}\label{refdefn}
So we now envisage a number $N$ of different coupling constants;
and we recognize that we might have to allow extra terms for interactions, in particular 
those whose bare couplings are zero (at least in the limit of greatest interest, viz. $d 
\raw 0$). This suggests a more sophisticated, indeed more flexible, task than I stated 
before (cf. after eq. \ref{gmu} in Section \ref{cut}). The task is still to assign bare 
constants so as to recover the measured physical constants, and in particular so as to 
secure the limit in eq \ref{limg0}. But now we are allowed to add (if need be) extra terms: 
terms which can be judiciously selected by us the theorist.

\indent It seems reasonable to say that such extra terms are  legitimate hypotheses to add 
to our initial theory (our initial collection of terms), {\em provided that} all the terms 
taken together, together with the limiting values of the bare constants given by eq 
\ref{limg0}, imply the measured values of the various $g(\mu)$. After all: we have at least 
`saved the phenomena' with our theory formulated on a spacetime continuum, albeit perhaps 
with the cost of judiciously selected extra terms. And this seems legitimate, even if (as 
conceded in Section \ref{dto0}) some of the
limiting values of the bare constants are $\infty$. Indeed: this seems legitimate, even if 
some of the limiting values of the bare constants in the new additional terms selected by us 
theorists are $\infty$---even though we  originally motivated such terms by the case where 
the limiting value of the additional bare constant is zero.\footnote{By focussing on 
renormalization, I have of course set aside the other requirements that the theory must 
satisfy, if we are to talk of `legitimate hypotheses' and `saving the phenomena'. To include 
those requirements, I should of course add something like: `and provided that the theory is 
otherwise empirically successful'.}

 In any case, whether or not you would call it `reasonable': this is the consensus, on the 
traditional approach to renormalization,
under one proviso. Namely, that there should be  only a {\em finite} number of extra terms. 
The idea is: our theory should not qualify as `saving the phenomena' if we have to make 
infinitely many such augmentations to it. That is: a theory which secures the limit in eq 
\ref{limg0}, using either no extra terms, or only a  finite number of them, is given the 
honorific adjective: {\em renormalizable}.


\subsection{Which theories are renormalizable?}\label{dyson}
I end this Section's review of the traditional approach to renormalization by very briefly 
reporting: (i) the criterion for when a theory is renormalizable, and (ii) that our 
empirically successful quantum field theories satisfy this criterion. The good fortune in 
(ii) will prompt the question: why should we be so lucky? Section \ref{modnut} will take up 
this question (using the criterion in (i)).

\subsubsection{Dyson's criterion}\label{dyson1}
Suppose we focus, not on a whole theory as given by a Lagrangian or Hamiltonian, i.e. by a 
sum of terms for the various sorts of energy of the various fields and their various 
interactions; but on a single such term. If you like, we imagine a theory so simple as to 
contain only one term. It turns out that the criterion for this theory, i.e. term, to be 
renormalizable, can be simply stated.

To do so, we should first express all dimensions in terms of length. We saw in Section 
\ref{cut} that, thanks to de Broglie's relation $p= h/{\lambda}$, we can trade in a cut-off 
in energy $\Lambda$ for a  distance $d$, and similarly the energy-scale $\mu$ for a distance 
$L$; with higher energies corresponding to shorter distances, so that e.g. $\mu \sim 1/L$. 
(Recall that $L$ is not the cut-off $d$.) But it turns out that we can go much further: not 
only energy but all quantities can be expressed as powers of length, by invoking the idea 
(due to Planck) that two fundamental constants, such as the speed of light $c$ and Planck's 
constant $h$, can be declared to be {\em dimensionless}, instead of (as usual) thinking of 
them as having respectively dimensions `length divided by time' and `length times momentum'. 
The idea is that after making this declaration, we `back-calculate' what dimension some 
familiar quantity such as an electric charge must have, so that our equations come out as 
dimensionally consistent. In this sort of way, a quantity turns out to have as its dimension 
some power of length: it has dimension length$^D$. Here, the {\em power} (also called: {\em 
exponent}) $D$ can be positive or negative. For example, $L^{-1} \equiv 1/L$, so that with 
$h$ declared dimensionless, de Broglie's relation $p= h/{\lambda}$ implies that momentum has 
dimension length$^{-1}$. For brevity, this is often shortened by dropping mention of length, 
so that we say: `momentum has dimension -1'.

We can now state the criterion for  a term (in the Lagrangian) to be renormalizable. It 
turns out that this is so iff: the bare coupling constant which the term contains has 
dimensions of length$^D$, with $D \leq 0$. This is called {\em Dyson's criterion}, or the 
{\em power-counting criterion}.

More precisely: suppose  that the bare coupling constant $g_0$ has dimensions of length$^D$. 
Then the corresponding physical coupling constant $g(\mu) \equiv g(L)$ will scale roughly 
like $L^{-D}$. That is:
\be\label{gLscale}
g(L)/g_0 \sim (L/d)^{-D}
\ee
Thus if $D > 0$, the exponent on the right-hand side will be negative; so when $L$ is very 
small, i.e. much smaller than $d$, the right hand side is very large. That is: the physical 
coupling constant will be large compared with the bare one. That is a sign of bad  behaviour 
at small distances $L$, i.e. high energies. At least, it is bad in the sense that the large 
coupling constant will prevent our treating the interaction represented by the term as a 
small perturbation. Thus  it is perhaps unsurprising that such a term is non-renormalizable 
in the sense sketched informally  at the end of Section \ref{nonren}.\footnote{This bad 
behaviour is not to say that a non-renormalizable theory is mathematically inconsistent: 
e.g. the Gross-Neveu model is non-renormalizable.}

Eq. \ref{gLscale} will also be important in the modern approach to renormalization. To anticipate a little: Section \ref{dwindle} will examine the case $D > 0$, i.e. non-renormalizability, for large distances; $L$ and Section \ref{beta} will examine the ``happy'' case of $D \leq 0$, even of small $L$.

\subsubsection{Our good fortune}\label{dyson2}
So much for the general ideas. How do the quantum field theories we ``believe in'', or 
``take seriously'' fare? That is: are the theories which are our best descriptions of the 
electromagnetic, weak and strong forces, renormalizable in the sense just discussed?

In short, they fare very well. For first: quantum electrodynamics (QED) is renormalizable in 
this Dyson sense. As to the other two forces: we have since the 1970s had:\\
\indent (i) a unified theory of the electromagnetic and weak forces (the {\em electro-weak 
theory} of Weinberg and Salam; also called `(quantum) flavour-dynamics' (QFD); and \\
\indent (ii) a theory of the strong force (quantum chromodynamics, QCD).\\
(Like QED, these theories are so far defined only perturbatively; but unlike QED, they each 
use a non-abelian gauge group: QFD uses $SU(2) \times U(1)$ and QCD uses $SU(3)$.) And 
indeed: both of these are renormalizable.

So all three---QED, QFD and QCD---are renormalizable. But we should recall that all three 
theories are defined only perturbatively: recall that we do not have a rigorously defined 
interacting quantum field theory in four spacetime dimensions (Section \ref{vpptbn}), and 
that even a finite theory is defined only perturbatively and may harbour divergent series 
(Section \ref{dto0}). Because of these limitations, a more modest jargon is appropriate. So 
a qualifying adverb is often added to the honorific `renormalizable'. Namely, we say these 
three theories are {\em perturbatively/superficially renormalizable}.

It seems a piece of great good fortune that our best theory of some force of nature be 
renormalizable (even perturbatively): let alone our theories of three such forces. At least, 
it is a relief after (a) having to admit that we can so far only define the theory 
perturbatively, and (b) having to face, from Section \ref{tradnutprosp} onwards, complicated 
corrections: a relief that the theory can in the above sense `save the phenomena', even if 
it is not finite in Section \ref{dto0}'s sense.

But we will now see that according to the modern approach to renormalization, this great 
good fortune is not so surprising. In a certain sense, {\em renormalizability is generic} at 
the low-ish energy scales we can access---cf. the next Section.

\section{The modern approach to renormalization}\label{modnut}
The key initial idea of this approach, initiated in the mid-1960s by the work of Wilson, 
Fisher, Kadanoff and others (with important earlier work by e.g. Stueckelberg, Gell-Mann and 
Low) is that instead of being concerned with good limiting behaviour as the cut-off $d \raw 
0$, we instead focus on how $g(\mu)$ varies with $\mu$. In terms of the ping pong ball analogy at the start of Section \ref{rencc}, and Section \ref{scal}'s discussion of energy scales: we now focus, not on regularizing integrals with a cut-off $d$, but on how the parameters of a system, e.g. the mass of a ping pong ball, depend on the energy or momentum scale at which we describe it.

Indeed, if we envisage a number of coupling constants, say $N$ for $N$ possible 
interactions, then the ``vector'' of coupling constants $(g_1(\mu),..., g_N(\mu))$ 
represents a point in an $N$-dimensional space; and as $\mu$ varies, this point flows 
through the space.   And accordingly: if we envisage a theory as given by a Lagrangian (or 
Hamiltonian) which is a sum of terms representing different possible interactions, then this 
space is a space of theories. Jargon: we say the coupling constants {\em run}, and the flow 
is called the {\em renormalization group flow.}

As we shall see, this simple idea leads to a powerful framework. I shall first (Section 
\ref{dwindle}) report how it explains why a theory (like QED) that concerns phenomena at the 
low (or low-ish!) energy scales that we can access, is renormalizable. That is: it explains 
why the good fortune noted in Section \ref{dyson2} is generic. Then in Section \ref{beta}, I 
discuss high-energy, i.e. short-distance, behaviour. Finally, I discuss insights about the 
renormalization group that come from thinking about statistical mechanics (Section 
\ref{smperspec}). All three Subsections will introduce some jargon, indeed ``buzz-words'', 
such as (respectively): fixed points,  asymptotic freedom and universality.

\subsection{Good fortune explained: non-renormalizable terms dwindle at longer 
distances}\label{dwindle}
To explain ``our good fortune'' in the sense introduced in Section \ref{dyson2} is to 
explain why a theory about phenomena at the low, or moderate, energy scales that we can 
access should be renormalizable. There are of course various philosophical controversies 
about explanation. But I take it to be uncontroversial that one very satisfying way to 
explain this would be to show: not merely that some given theory is renormalizable; but that 
{\em any} theory, or more modestly, any of a large and-or generic class of theories, is 
renormalizable.  To the extent that the class of theories is indeed large and-or generic, 
such an argument would demonstrate that our good fortune was ``to be expected''. 
(Admittedly, such an explanation, whether for a single theory, or for a class of them, will  
have to make some other assumptions about the theory or theories: a point I will stress in 
Section \ref{decouple}. So it is only relative to those assumptions that the good fortune is 
explained, and to be expected.)

This is indeed what the modern approach to renormalization gives us, with its idea of a 
space  of theories, on which there is a flow given by varying the energy-scale $\mu$. More 
precisely and modestly, but also more practically: I admit that this approach does not show 
that any of a large and-or generic class of theories has, at the comparatively low energies 
and large length-scales we can access, literally {\em no} non-renormalizable terms. Rather, 
the approach shows that for any such theory---``with whatever high-energy behaviour, e.g. 
non-renormalizable terms, you like''---the non-renormalizable terms dwindle into 
insignificance as  energies become lower and length-scales larger. That is, in Section 
\ref{tradnut}'s notation: the physical coupling constant for non-renormalizable terms 
shrinks. For such terms: as $\mu \raw 0$ (i.e. $L \raw \infty$), $g(\mu) \equiv g(L)$ $\raw 
0$.

Indeed, this explanation is already clear from Section \ref{dyson1}'s discussion of eq. 
\ref{gLscale}: which, to repeat it, was:
\be\label{gLscalerepeat}
g(L)/g_0 \sim (L/d)^{-D}
\ee
In Section \ref{dyson1}, we focussed on the case where $L$ is very small, so that a 
non-renormalizable term's positive exponent (in the dimension of length) makes for a large 
physical coupling constant. {\em But just look at the other side of the same coin.} When $L$ 
is large (much larger than the cut-off $d$), and $D > 0$ (i.e. the term in question is 
non-renormalizable), then the right hand side of eq. \ref{gLscalerepeat} is very small. That 
is: the physical coupling constant is very small. So at large distances, the 
non-renormalizable interaction is weak: ``you will not see it''.\\

There are four main points I should make about this explanation, before discussing 
short-distance behaviour (Section \ref{beta}). The first point is about how non-trivial the 
{\em explanans}, i.e. eq. \ref{gLscalerepeat}, is. The second point is about the explanation 
not depending on spacetime being a continuum.  This will prompt the third point, about {\em 
effective theories}. (The second and third points, in Sections \ref{notcontm} and \ref{eft?}, correspond to Sections 5.2.1 and 5.2.2 of the companion paper (2014).) The fourth point will somewhat generalize the discussion, from a physical not philosophical viewpoint; and will introduce some more jargon.

\subsubsection{Decoupling high-energy behaviour}\label{decouple}
That at large distances, a non-renormalizable interaction is weak follows immediately from 
eq. \ref{gLscalerepeat}. But that does {\em not} make it obvious! A good deal of theory 
needs to be assumed in order to deduce eq. \ref{gLscalerepeat}. After all, there is of 
course no {\em a priori} guarantee that interactions that are strong at short distances 
should be weak at long distances. To show this ``decoupling'' of high-energy behaviour from 
the low-energy behaviour was a major achievement of Wilson, the other authors mentioned, and 
indeed many other physicists, e.g. Symanzik (1973), Applequist and Carazzone (1975). I will 
not go into details, but just make three general  remarks.\\
\indent (i): It can be shown under very general conditions, even within the confines of a 
perturbative analysis. \\
\indent (ii): Looking ahead: Section \ref{eft?} will mention Weinberg's perspective, based 
on a result roughly to the effect that, even without assuming the framework of quantum field 
theory {\em ab initio}, any relativistic quantum theory's description of physics at low 
enough energies must look like the description given by a quantum field theory.\\
\indent (iii): Again, looking ahead: Section \ref{beta} will say a bit more about how the 
limiting high-energy behaviour of $g(\mu)$ is encoded in a function, the beta-function, 
which can be calculated perturbatively.

\subsubsection{Spacetime need not be a continuum}\label{notcontm}
Notice that this explanation does not depend on our theory (with all its terms, including 
non-renormalizable ones) being true, or even approximately true, at arbitrarily short 
distances. It only needs to be approximately true at suitable intermediate distances.  More 
precisely: it only needs to secure eq. \ref{gLscalerepeat} holding for any  
non-renormalizable interaction at a range of scales which is wide enough to include $L$  
being sufficiently larger than the cut-off $d$, so that with the given positive dimension  
$D$ of the bare coupling constant, the left hand side of eq. \ref{gLscalerepeat} is small  
enough that we will not see the interaction.

We can put the same point in more physical terms, and in terms of energies. Maybe at very 
high energies, spacetime does not behave like a continuum.  But provided the theory is 
``true enough'' at some high, maybe even inaccessible, energies in the sense that it 
validates eq. \ref{gLscalerepeat}, then we can deduce that at much lower, in particular 
accessible,  
energies, ``we see only renormalizable interactions''. That is: our theory's predictions  
have significant contributions only from renormalizable interactions.

Note incidentally that this independence of spacetime being a continuum is much stronger 
than we saw before, in the shift in Section \ref{dto0}, from assuming (a) and (b) to 
assuming (a') and (b'): roughly, the shift from assuming that our theory described physics 
in a continuous spacetime to assuming merely that it might do so. In the present argument, 
we could be agnostic about whether, or even deny that, our theory could describe physics in 
a continuous spacetime. All we need is that it is approximately true at suitable 
intermediate distances, as just specified.

Here we meet a widespread jargon. A theory that is taken to be approximately true in a given 
regime (of energy and-or length, and-or some other parameters) is called {\em effective}. 
The adjective is  used especially when the theory is known or believed to be {\em only} 
approximately true; say, because it is derived using certain approximating and-or idealizing 
assumptions (assumptions which go beyond merely specifying the regime, i.e. range of 
parameters, concerned).

So we can sum up the above explanation of what I called `our good fortune' by saying: from 
studying the renormalization group flow, we deduce (subject to the theoretical assumptions 
gestured at in Section \ref{decouple})  that effective low-energy theories are 
renormalizable. The idea of effective theories leads in to the next point.

\subsubsection{Effective theories only?}\label{eft?}
I ended Section \ref{dto0} by reporting that on the traditional approach, renormalizability 
functioned as a criterion for selecting theories. But the  explanation at the start of 
Section \ref{dwindle} undermines this stance. For it says that, although non-renormalizable 
terms induce bad behaviour, i.e. a large coupling, at short distances, this bad behavour is 
invisible at the larger distances we can access. So why worry? That is: why not countenance 
non-renormalizable terms, at least for inaccessibly high energies?

Of course, the words `worry' and `countenance' are vague. What you are inclined to worry 
about, and correspondingly what you are willing to countenance, will depend on your 
background attitudes to quantum field theory: for example, on how confident you are about 
using it at high energies, and about accepting results obtained from a heuristic formalism, 
rather than by rigorous mathematical proofs. So there are bound to be several possible 
positions. Here I will briefly develop one position, often called the {\em effective field 
theory programme} (or: approach). It is based, not so much on confidence about the two topics above, as on an opportunistic or instrumentalist attitude to being {\em unconfident} about 
them. (In Section \ref{beta}, I will describe a less opportunistic or instrumentalist 
attitude, based on results showing some quantum field theories' good behaviour at 
arbitrarily short distances.)

There are of course two main factors that prompt a cautious or sceptical attitude towards 
the framework  of quantum field theory.\\
\indent (1): One is just that interacting quantum field theories (in four spacetime 
dimensions) are at present mathematically ill-defined. Recall that this was the first of the 
two serious mathematical problems listed in Section \ref{vpptbn}.\\
\indent (2): The other factor is the expectation that at sufficiently high energies, the 
framework breaks down, to be replaced by a theory or theories using a different framework. 
This break-down might occur only at the vast energies
associated with quantum gravity: the replacement theory being perhaps a version of string 
theory, or some other current contender for a theory of quantum gravity. Or the break-down 
might occur at intermediate energies, energies far higher than we can (and probably: ever 
will) access, but well below those of quantum gravity: there are proposals for new 
frameworks at these energies, such as non-commutative geometry.

Either or both of these factors prompt one to be cautious about drawing from quantum field 
theory conclusions about ontology. Or rather: conclusions about the ontology of phenomena at 
very high energies, or very short distances. But these factors should not suspend all 
discussion of ontology in the light of physics, or even in the light of quantum field 
theory; for three reasons. \\
\indent (a): Whatever the phenomena at very high energies turn out to be, whatever the 
theoretical framework for describing them, and whatever ontology that framework suggests, we 
have every reason to expect that the facts at those energies determine (in philosophical 
jargon: {\em subvene}) the facts at the lower energies we can access. \\
\indent (b): And given the great success of quantum field theory, we have every reason to 
expect that the facts at those very high energies imply a quantum field theoretic 
description at the lower, accessible, energies. \\
\indent (c): Besides, whoever said that ontology concerns only ``the supervenience basis'', 
i.e. the putative set or level of facts that determine (subvene) all other facts? That is: 
there is plenty of scope for ontological discussion of supervening (``higher level'') facts 
and theories: in particular, there is scope for for ontological discussion of quantum field 
theory.

But these factors also suggest that even below the energy scale at which the entire 
framework of quantum field theory breaks down, there may, for all we know, {\em not} be any 
single quantum field theory which is more fundamental than the others, in the sense that 
each of them is derived from it by assuming extra conditions that specify the derived 
theory's regime (of energies and types of interaction considered etc.). That is: as the 
energy scale gets higher and higher (while remaining below the scale at which the entire 
framework of quantum field theory breaks down), physics might be described by a succession 
of quantum field theories, each of which accurately describes the phenomena at a certain 
range of energies, but becomes inaccurate above that range. And when it becomes inaccurate, 
it may also become even more badly behaved, mathematically.

This scenario is often called the {\em tower of effective field theories}. But the phrase 
can be misleading, for two complementary reasons. \\
\indent (i): First, as I mentioned when defining `effective', at the end of Section 
\ref{notcontm}: the adjective is often used when the theory is known or believed to be only 
approximately true, because it is derived using  approximating and-or idealizing 
assumptions. But in this scenario, the theories in the envisaged tower are {\em not} 
required to be derivable from some other theory: in particular, one with greater 
credentials, or warrant, for being taken as exactly true (`fundamental') because it also 
covers higher energies. Rather, each theory  is simply accurate in its energy range, and 
inaccurate beyond it. \\
\indent (ii): Second: the word `tower' suggests an infinite tower. But as I noted in (1) 
above, there are good reasons (concerning  quantum gravity if nothing else) to think that at 
{\em some} energy, the entire framework of quantum field theory breaks down. So as a 
programme or approach for quantum field theory, the effective field theory programme can, 
and should, admit that the tower is probably finite.

But setting aside misleading connotations: the main point is that this scenario gets some 
support from this Section's explanation of ``our good fortune'', viz. that any 
non-renormalizable interactions (terms), though they would be important at higher energies, 
will make a dwindling contribution to all processes, as the energy scale is reduced. For 
this explanation implies that we cannot get evidence about which non-renormalizable 
interactions, if any, operate at inaccessibly high energies. Whatever they are---and 
whatever the bad short-distance behaviour they suffer (cf. the end of Section 
\ref{dyson1})---we will not see them. So why worry about non-renormalizable interactions 
(terms)? And for all we know, or could ever know, the scenario of the tower holds good: 
there is no fundamental quantum field theory, and various such interactions operate at 
various inaccessibly high energies.

There is a further, and substantial, point to make. So far, my exposition of the effective 
field theory scenario has had the spirit of epistemic modesty: ``for all we know''. A true 
and worthy sentiment, if a bit dull. But Weinberg has developed a stronger and more positive 
perspective on the matter. It provides an answer to the question why physics at accessible 
energies should be described by a quantum field theory at all, {\em even if} the framework 
breaks down higher up, e.g. because of gravity. And this answer yields the narrative 
strategy for his magisterial exposition of quantum field theory (1995; cf. pp. xx-xxi, 1-2, 
31-38; 1999 pp. 242-247). In short, there is the following result; (admittedly, with 
`result' understood by the standards of heuristic quantum field theory, not pure 
mathematics). Any quantum theory that at low enough energies is Lorentz-invariant and 
satisfies one other main assumption, called `cluster decomposition' (which is plausible, 
since it has the flavour of a locality assumption), must at low enough energies be a quantum 
field theory (1999, p. 246).

So much by way of sketching the effective field theory programme. We can sum it up as urging 
that, regardless of how and why quantum field theory might break down at very high energies 
(as it presumably does, thanks to gravity, if nothing else): we have no reason in theory, 
nor experimental data, to deny the scenario of the tower---a succession of theories, each 
accurately describing physics in its energy range, and inaccurate beyond it.

As I said at the start of this Subsection, this suggests a rather opportunistic or 
instrumentalist attitude to quantum field theories. I will return to this briefly at the start of  Section \ref{Nag}.  Meanwhile, in Section \ref{beta}, I will describe how results showing some 
quantum field theories' {\em good} behaviour at arbitrarily high energies foster a less 
opportunistic or instrumentalist attitude. More precisely: the results suggest that there 
are good prospects that these theories can be rigorously defined ({\em pace} (1) above).

\subsubsection{The renormalization group flow}\label{flow}
So far, my talk of the renormalization group flow has been restricted in three ways; which 
we need to overcome. The most important is that we  need to consider, not just the flow as 
energy $\mu$ decreases (length $L$ increases), but also the flow in the other direction: as 
$\mu$ increases ($L$ decreases). This needs a separate subsection: Section \ref{beta}. Here 
I overcome two smaller restrictions:  this will also introduce more jargon.

\indent \indent  (a): A flow can have a {\em fixed point}, i.e. a point that is not moved by 
the flow: think of sources and sinks in elementary discussions of fluid flow. In our context 
(the renormalization group flow), this would mean a set of physical coupling constants 
$(g_1(\mu),..., g_N(\mu))$ that is unchanged as $\mu$ decreases further (as the length-scale 
increases further). Jargon: the behaviour of the system is {\em scale-invariant}: ``you see 
the same behaviour/theory/physical coupling constants, at many different length-scales''. 
This can indeed happen; and we will see a vivid physical reason for this, related to 
statistical mechanics, in Section \ref{smperspec}. Such a point is called an {\em infra-red 
fixed point}. Here, `infra-red' is used on analogy with light: infra-red light has a longer 
wavelength, lower frequency and lower energy, than visible light.

\indent  \indent  (b):  So far, we have had in mind one trajectory, maybe leading to a fixed 
point. But many trajectories might lead to the same fixed point; or at least enter and 
remain in the same small region of the space. If so, then the  `vectors' $(g_1(\mu),..., 
g_N(\mu))$ at diverse early points on a pair of such trajectories representing dissimilar 
theories lead, as $\mu$ decreases, to the same fixed point, or at least to the same small 
region, and so to similar theories. That is: when you probe at low energies/long distances, 
``you see the same physical coupling constants/behaviour/theory''. Jargon: This is called 
{\em universality}. And the set of `vectors' that eventually lead, as $\mu$ decreases, to 
the given fixed point is called, on analogy with elementary discussions of fluid flow, the 
point's {\em basin of attraction}.  But note that universality should really be called 
`commonality' or `similarity': for there can be different fixed points, each with their own 
basin of attraction. But jargon aside: Section \ref{smperspec} will give more physical 
details about universality, and Section \ref{manna2} will assess whether it is different 
from the familiar philosophical idea of multiple realizability.

Finally, we can summarize this Subsection's main point, that non-renormalizable interactions 
dwindle at large length-scales, by combining the jargon we have just introduced with the 
previous jargon that a {\em free} theory is a theory with no interactions. Namely: the 
infra-red fixed points of theories all of whose terms are nonrenormalizable are free 
theories.

\subsection{Short-distance behaviour: the beta-function and asymptotic freedom}\label{beta}
Instead of considering the flow as energy $\mu$ decreases (length $L$ increases), we can of 
course consider flowing in the other direction: as $\mu$ increases ($L$ decreases). Again, 
the jargon is borrowed from light: we can consider the flow towards {\em  the ultra-violet}. 
Looking again at eq. \ref{gLscalerepeat} (which is eq. \ref{gLscale}), we see that it is 
terms/interactions for which $D < 0$ for which the physical coupling constant goes to zero 
as $L \raw 0$; since for these terms, the physical coupling constant scales like $L$ to a 
positive power.

 Of course, the coupling constant being zero means the interaction is not ``seen'' (cf. 
Section \ref{physrat}). The behaviour we see is that of the free, i.e. non-interacting, 
theory. This is called {\em asymptotic freedom}. And as in (a) in Section \ref{flow}, this 
free theory may be fixed, i.e. not moved, by the flow. If so, it is an {\em ultra-violet 
fixed point}.

On the other hand, if $D = 0$, then according to eq. \ref{gLscalerepeat}, the physical 
coupling constant scales  like $L$ to the zeroth power; that is, it is constant. More 
precisely: we need to study the range under which eq. \ref{gLscalerepeat}, or some more 
precise equation, is valid, and what happens to the physical coupling constant(s) beyond 
that range.

So these cases, $D < 0$ and $D = 0$, are very different; accordingly, there is jargon to 
distinguish them. Recall that in Section \ref{dyson1}, we called a term for which $D \leq 0$ 
`renormalizable'. But we now distinguish the two cases. If $D < 0$ (the ``happy case''), we 
say the theory is {\em super-renormalizable}.  If $D = 0$, we say the theory is (``merely'') 
renormalizable. But if in this latter case, a more subtle analysis shows that the coupling 
constant goes to zero as $L \raw 0$, we will still say the theory is {\em asymptotically 
free}.  That is: this buzz-word is not reserved for super-renormalizable theories.

We can summarize, using the jargon we have just introduced, like we did at the end of 
Section \ref{flow}. Namely: asymptotically free theories, in particular super-renormalizable 
theories, have a free theory as an ultra-violet fixed point.

Note that the idea of a ultra-violet fixed point is more general than asymptotic freedom, in 
that the renormalization group flow could have a non-free theory as an ultra-violet fixed 
point. The various possibilities for what happens to $g(\mu)$ as $\mu$ tends to infinity are 
often described in terms of the {\em beta-function}, which is defined by
\be
\beta (g) :=  \frac{d  g}{ d \ln \mu} \equiv \mu \frac {d g}{d \mu}.
\ee
Here $\ln \mu$ is the logarithm of $\mu$. So the $\beta$-function is the rate of increase of 
$g$ with respect to a logarithmically damped measure $\ln \mu$ of the energy: since 
logarithm rises very slowly, it is in effect an amplified rate of increase of $g$ with 
respect energy---amplified by multiplying by the energy itself.

So as $\mu$ tends to infinity, there are three possibilities for $g(\mu)$ having a finite 
limit, i.e. an ultra-violet fixed point. Since there is a fixed point, all will involve 
$\lim_{\mu \rightarrow \infty} \beta = 0$. But $g$ might have zero as its limit (as 
discussed: asymptotic freedom).  Or $g$ might have some non-zero value $g_\star$ as its 
limit. This is called {\em asymptotic safety}. Or $g$ might be a constant, $g_\star$ say, 
independent of $\mu$; so that $g$ does not, colloquially speaking, tend to a limit---it is 
already there. This situation occurs in theories which are {\em conformally 
invariant}.\footnote{This means, roughly speaking, that the theory is symmetric under any 
change of scale (a dilation). This extra symmetry makes conformally invariant theories 
easier to study in various ways (especially if spacetime has only one spatial dimension); 
and thus important to us, even though  they do not describe the known forces.}

To summarize, the three possibilities for $g(\mu)$ having a finite limit at short distances 
are:\\
 \indent(a): {\em asymptotic freedom}: $\lim_{\mu \rightarrow \infty} \beta = 0$;  
$\lim_{\mu \rightarrow \infty} g = 0$;\\
   \indent (b): {\em asymptotic safety}: $\lim_{\mu \rightarrow \infty} \beta = 0$;  
$\lim_{\mu \rightarrow \infty} g = g_\star  \neq 0$. \\
     \indent (c): {\em conformal  invariance}: $\beta \equiv 0$ i.e. $g$ is constant, 
independent of $\mu$.\\
Compare Figure 1; where the ultra-violet fixed point is the dot. In Fig. 1(a), as $\mu$ 
grows the negative $\beta$ drives $g$ down to 0. In Fig. 1(b), as $\mu$ grows, a positive 
$\beta$ drives $g$ up towards the fixed point, while a negative $\beta$ drives it down. 
Finally in Fig. 1(c), $g$ is constant independent of $\mu$.\footnote{My thanks to Nazim Bouatta for Figure 1, and for teaching me most of this Section. Of course, one can give 
a more fine-grained classification than Figure 1's (a)-(c): cf. e.g. the options in Weinberg (1995a, 
Section 18.3, p. 130f.).}

\begin{figure}[here]
\centerline{\includegraphics[width=7cm]{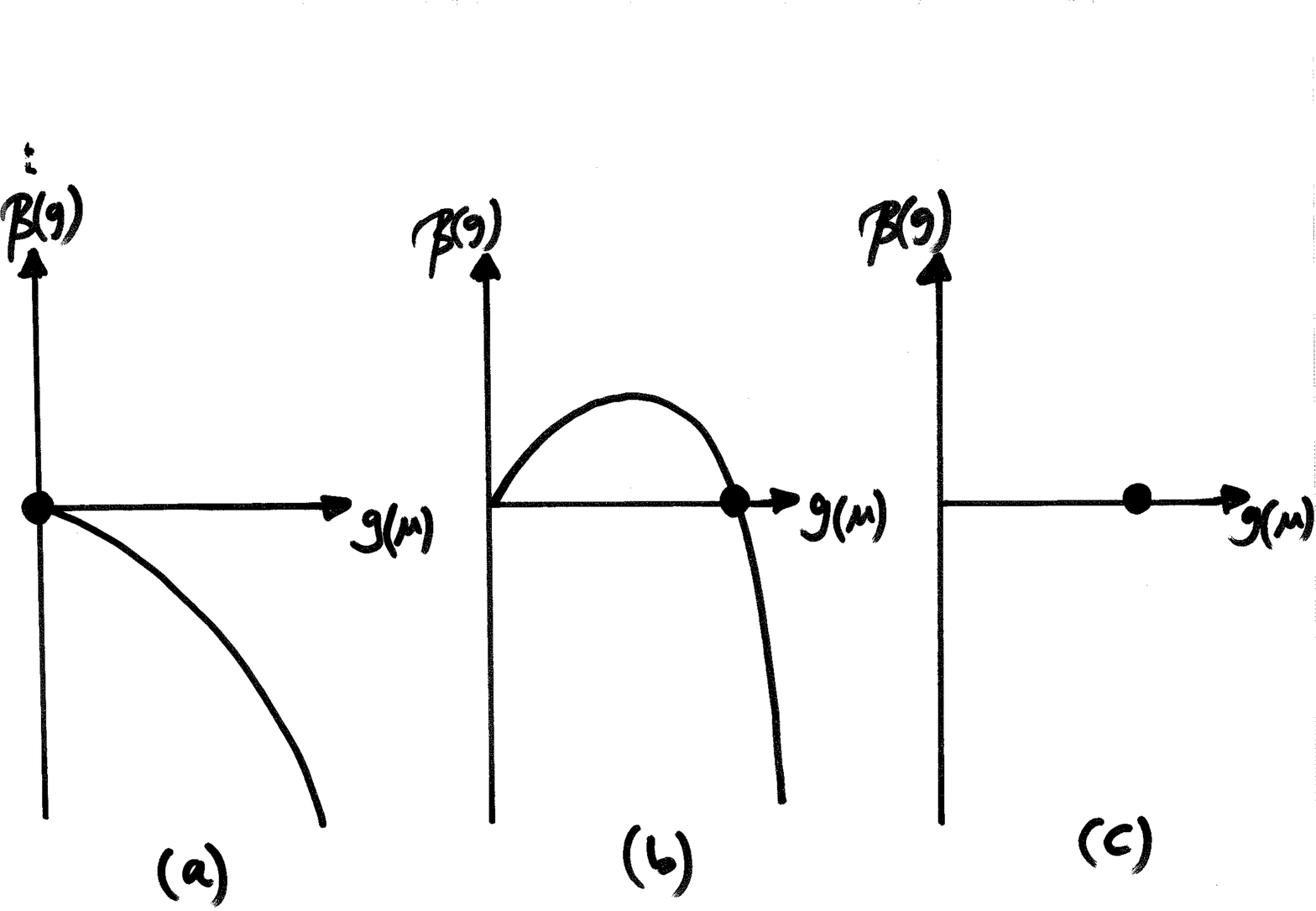}}\caption{UV fixed points}
\end{figure}

So much for the general ideas. Let us ask, as we did in Section \ref{dyson2}: How do the 
quantum field theories which are our best descriptions of the electromagnetic, weak and 
strong forces, get classified? There, we reported  that all three (i.e. QED, QFD and QCD) 
are renormalizable---in that Section's {\em inclusive} sense, that $D \leq 0$: i.e. $D < 0$ 
{\em or} $D = 0$. More exactly, they are perturbatively renormalizable, since as emphasized 
there, the theories have not yet been rigorously defined.

Now that we distinguish the two cases, $D < 0$ vs. $D = 0$, there is: bad news and good 
news---indeed, there are two pieces of each. First, the bad news: (that is, in addition to 
the prevailing lack of rigour). First: None of the theories is super-renormalizable. They 
are ``merely''  renormalizable; so we need a more subtle analysis of their short-distance 
behaviour. Because the three theories are only defined perturbatively, it is hard to make 
such an analysis. But there is every reason to believe that for QED, there is more bad news; 
(this is the second piece of bad news). Namely: QED's is badly behaved at short distances. 
That is: in QED, as $L$ decreases, the coupling constant, i.e. the charge of the electron, 
at first looks constant---but it then grows and grows. There is good reason to think it 
tends to infinity, as $L \raw 0$. 


On the other hand: for QCD, the corresponding analysis yields good news---good 
short-distance behaviour. That is: There is every reason to believe that QCD is 
asymptotically free. So at very short distances, quarks do not feel each other's weak or 
strong force.\footnote{Wilczek's Nobel lecture (2005) is a beautiful and masterly 
introduction to asymptotic freedom, especially in QCD. QFD is, unfortunately, not 
asymptotically free. Its high energy behaviour is complicated: for details cf. e.g. Horejsi 
(1996),  Moffat (2010); thanks to Nic Teh for these references.} Besides, there may be some 
good news about gravity. In this paper, I have of course ignored gravity, apart from saying 
in Section \ref{eft?} that people expect quantum gravity to force a breakdown of quantum 
field theory. One main reason for that is the fact that the quantum field theory obtained by 
quantizing general relativity is not renormalizable: and thereby, on Section \ref{tradnut}'s 
traditional approach, not acceptable. But there is evidence that this theory is 
asymptotically safe, i.e. that the physical coupling constant has a finite limit at high 
energies, case (b) above; (Weinberg 1979, Section 3, p. 798f).

This good news prompts a broader point, which was foreshadowed at the end of Section 
\ref{eft?}'s discussion of effective theories. Namely, asymptotic freedom suggests these 
theories {\em can} be rigorously defined. This is not to suggest that success is over the 
next hill: if attainable, it is some way off---but asymptotic freedom gives us grounds for 
optimism.\footnote{Besides, we can show that if  a theory rigorously exists, then its 
asymptotic freedom can be ascertained perturbatively: so there is no threat of future 
success undermining our present grounds for optimism. For this pleasant surprise, cf. Gross 
(1999, p. 571).} If so, this would count against the effective field theory vision, that 
(regardless of gravity) there is a succession of theories, each accurately describing 
physics in its energy range, but inaccurate beyond it.

\subsection{The perspective from the theory of condensed matter}\label{smperspec}
No account of the modern approach to renormalization, however brief, would be complete 
without some discussion of the role therein of ideas from the theory of condensed matter. 
(`Condensed matter' is short for `liquid or solid'.) Ideas from this field have been 
invaluable. To convey something of these insights, I shall make just three main points: that 
continuous phase transitions correspond to infra-red fixed points of the renormalization 
group flow (Section \ref {crpt}); that renormalization group methods enable us to calculate 
correctly the critical exponents of such transitions (Section \ref {cret}); and finally, 
that in a condensed matter system, there is a natural lower limit to the cut-off $d$ and 
length $L$ (Section \ref {latt}).\footnote{Condensed matter is, fortunately, more familiar 
than quantum fields. Among many approachable references for this material, let me pick out 
just Kadanoff's masterly surveys (2009, 2013),  Batterman (2010), and Menon and Callender 
(2013).}

At the outset, I should clarify my usage. I will contrast `theory of condensed matter', with 
`quantum field theory', understood as I have done hitherto in this paper: viz. as describing 
high energy physics, especially the fundamental forces---electromagnetism, the weak force 
and the strong force. But I stress that the mathematics of quantum field theory is used 
endemically to describe condensed matter. For example, one often describes a solid or liquid 
with a quantum field (say: energy or momentum, or electric field): this amounts to assigning 
a quantum operator to each point of space or spacetime---thus abstracting away from the 
atomic constitution of matter. I will briefly return to this in Section \ref{latt}.

\subsubsection{Continuous phase transitions: scale-invariance}\label{crpt}
In both classical and quantum physics, understanding  condensed matter is usually harder 
than understanding gases, since the mutual proximity of the atoms makes for stronger 
interactions, and so for problems that are harder to solve (cf. Section \ref{vpptbn}'s 
discussion of intractable Hamiltonians). So it is little wonder that most of the early 
successes of statistical mechanics---which is the core theory for understanding condensed 
matter---concerned gases. But  the last half-century has seen great strides in understanding 
liquids and solids, in both classical and quantum physics.

Most relevant to us is the topic of {\em phase transitions}. These are transitions between 
macroscopically distinguishable states (also called: {\em phases}), such as the liquid, 
solid or gaseous states themselves. So melting and freezing, boiling and condensing, are all 
phase transitions; but so is the change in the direction of magnetization of some material, 
under the influence of a changing magnetic field. Here, I will consider only a special kind 
of phase transition, called {\em continuous} or {\em second-order}, or a {\em critical 
point}. The idea is that in such a phase transition, the material ``looks the same'' at 
whichever length scale you examine it ({\em scale-invariance}): this phenomenon of 
scale-invariance does not occur in the more usual {\em first-order} phase transitions.

 Thus consider water boiling, i.e. liquid water being heated and becoming steam. 
Usually---for example in a kettle---the phase transition is first-order: there is a body of 
water that is almost entirely liquid but for a few bubbles of steam rising in it, and a body 
of steam (and air), but for a few droplets of liquid water. If we think of the density as 
the quantity whose varying values encode the difference between the phases, liquid and 
solid, there is no scale-invariance. For the two regions where liquid and gas predominate 
are separated by the bubbling surface; and besides, on a smaller scale, the bubbles of steam 
in the liquid (and the droplets of water in the gas) have some characteristic mean size (and 
variance).

But there is a special temperature and pressure, called the {\em critical point}, at which the distinction between liquid water and steam breaks down, and there {\em is} scale-invariance. That is: scrutinizing a 
volume of liquid water, we see that it contains bubbles of steam of widely varying sizes in 
roughly equal proportions; and scrutinizing such a bubble, we see that {\em it} contains yet 
smaller droplets of water, which are themselves of widely varying sizes in roughly equal 
proportions; and if we scrutinize one of those droplets, we find it contains yet smaller 
bubbles ... and so on, through many orders of magnitude of size, until we reach molecular 
dimensions, where, of course, the alternation between the phases breaks down.\footnote{This 
critical point for water happens at a temperature of 374 degrees Celsius (647 = 374 + 273 
degrees Kelvin). The water-steam mixture takes on a cloudy appearance so that images are 
blurred; and thus the phenomenon is called `critical opalescence'. As we will see in Section 
\ref{cret}, it also happens for other liquid-gas mixtures.}

Thus the critical point involves a ``tower of self-similarity'', where zooming in to see 
finer detail presents the same picture as we saw at larger length-scales. More precisely: it 
presents the same sort of picture, in a statistical sense. That is: the exact position, size 
and other properties of the bubbles of steam, at any level, is  of course a matter of 
myriadly complicated happenstance. But the statistical properties of the bubbles at 
different levels match, in the sense that: if we specify the degree of magnification 
(zooming in: the change of length-scale) between two levels, there  is a corresponding  
re-scaling of  the bubbles' other quantities, such as expected size, density etc., which 
maps the means and variances of bubbles' quantities at the first level to those at the 
second. In short: there is statistical self-similarity, under changes of length-scale, 
through many orders of magnitude, until we reach molecular dimensions.

Many other phase transitions, in many other materials, can occur in this curious, 
statistically self-similar, way in which the idea of a boundary between the phases breaks 
down; (unsurprisingly, this requires special external conditions, like temperatures, 
pressures etc.). For example, it can happen in a magnet. The analogue to the alternation 
between bubbles of steam and droplets of liquid water is the alternation between the 
magnetization in two different spatial directions, for example ``up'' and ``down''. At the 
critical point (requiring a special temperature, the Curie temperature), a region of the 
magnet with magnetization predominantly up turns out to contain ``islands'' whose 
magnetization is predominantly down, but each such island contains contain islands whose 
magnetization is predominantly up ... and so on.

We can already see how the idea of a critical point  connects with several of the notions in 
Section \ref{dwindle}, especially the renormalization group flow, infra-red fixed points and 
universality (Section \ref{flow}). Zooming out our description of a system to longer 
distances corresponds to flowing to lower energies (decreasing $\mu$) in a quantum field 
theory. Scale-invariance means that the description does not change as we zoom out further. 
So such a description  corresponds to an infra-red fixed point of the renormalization group 
flow.

Furthermore, we can strengthen my phrase `corresponds to' to {\em is}, if we make the notion of `description' of the condensed matter system more precise as the set of physical coupling 
constants that occur in the Hamiltonian that describes the system  at the distance-scale 
concerned. (Similarly, with Lagrangian instead of Hamiltonian; but for short, I will just 
say `Hamiltonian'.) 

That is: we can set up the idea of a space of theories of condensed 
matter systems, and a flow on this space, just as we did at the start of Section 
\ref{modnut}. Given a postulated microscopic Hamiltonian describing the atomic or molecular 
interactions, the zooming out process is then a matter of successively coarse-graining the 
description, i.e. defining collective variables, to give an effective Hamiltonian. The standard example, viz. the Ising model, is very intuitive. The system is a regular array of sites (in one or more dimensions: a line or a lattice); with each of which is associated a variable taking one of two values, +1 or -1. The microscopic Hamiltonian encodes the idea that equal values for neighbouring spins are ``preferred'' (i.e. have lower energy). One coarse-grains by taking a majority vote of the values in a block of, say, ten spins; thus defining a new array, with ten times fewer sites, described by a new effective Hamiltonian. One then defines a flow on the space of Hamiltonians by iterating this coarse-graining; for details, cf. the {\em maestro}'s exposition (Kadanoff 2013, Section 6.4, p. 170-172). Thus a Hamiltonian that is unchanged by zooming out, i.e. is scale-invariant, is precisely an infra-red fixed point of this flow.

Finally, note that the notion of universality carries over directly to the context of 
condensed matter physics. It means that two disparate physical systems, with very different 
microscopic Hamiltonians, can be described at long length-scales in the same way. (This of 
course  sounds exactly like philosophers' notion of multiple realizability: cf. Section 
\ref{manna2}.) Indeed: if the two microscopic Hamiltonians flow to the same infra-red fixed 
point, then the two systems are described at their respective critical points, by the very 
same Hamiltonian. (And similarly, for descriptions in the neighbourhood of the critical 
point: the descriptions of the two systems are close but not identical.) Having the same 
Hamiltonian can make the two systems have  exactly equal values for corresponding 
quantities. This quantitative equality, despite very disparate microscopic constitutions, 
can be very striking---as we will see in the next Subsection.

\subsubsection{Critical exponents: the correlation length}\label{cret}
One aspect of the quantitative behaviour of materials at and near a continuous phase 
transition, is the fact that the values  of various quantities are given by power-laws, i.e. 
by the value of  some other quantity, raised to some power. More specifically:  critical 
points occur only at a specific temperature, the {\em critical temperature} $T_c$; and near 
the critical point, the value of some quantity, $v(Q)$ say, is given  by a power of the 
difference between the actual temperature $T$ and $T_c$, or by a power of some similar 
quantity such as the ratio of this difference to $T_c$:
\be\label{cretgenl}
v(Q) \sim |T - T_c|^p \;\;\; {\mbox{or}} \;\;\; v(Q) \sim |\frac{T - T_c}{T_c}|^p
\ee
where $p$ is the power.

It is worth giving some examples of how the same power law (same $p$) can govern the 
critical points of very disparate systems. This will show how striking universality can be: 
which will be relevant to the philosophical discussion in Section \ref{manna2}. Besides, 
since the renormalization group framework can correctly predict these power laws for 
countless such systems, it will show the amazing success of the framework. My examples will 
be drawn from condensed matter, not quantum field theory, since this will be technically 
less demanding: water and steam are familiar, while quarks and gluons are not. But I stress 
that this Section's themes---phase transitions, critical points, universality and the 
renormalization group successfully predicting the power laws---nowadays also form a large 
subject within quantum field theories like QCD; (e.g. Kogut and Stephanov 2004).  But note: 
the details in the rest of this Subsection are {\em not} needed in my closing philosophical 
discussion (Section \ref{Nag}), and so can be skipped; nor will Section \ref{Nag} need  
these themes in the form they take within quantum field theories.

In eq. \ref{cretgenl}, $Q$ might be a quantity whose value  is zero at the critical point 
(so that $p$ is positive). For example, $Q$ might be:\\
\indent (i) the difference $\rho_{\rm{liquid}} - \rho_{\rm{gas}}$ between the densities of 
liquid water and of steam; or \\
\indent (ii) the average magnetization $m$ of a piece of iron.\\
Or $Q$ might be a quantity that diverges (i.e. goes to infinity) at the critical point (so 
that $p$ is negative). For example, $Q$ might be \\
\indent (i') the isothermal compressibility $\kappa_T$ of water: a measure of how 
compressible it is, i.e. how the volume changes as a function of pressure, at fixed 
temperature $T$ (formally: $\kappa_T = - {\partial V}/{\partial  p}|_T$); or \\
\indent (ii')  the magnetic susceptibility of iron: a measure of how magnetizable it is, 
i.e. how its magnetization changes as a function of an external applied magnetic field $B$, 
at fixed temperature $T$; (formally: $\chi_T = - {\partial m}/{\partial B}|_T$).

In these four examples, it is obvious enough that (i), the difference $\rho_{\rm{liquid}} - 
\rho_{\rm{gas}}$, vanishes at the critical point. For as I sketched in Section \ref{crpt}, 
each body of liquid water contains much gas, which contains much liquid, and so on ... and 
{\em vice versa}. A similar argument can be made that (ii) vanishes, and (i') and (ii') 
diverge, at their respective critical points. But we do not need the details of why that is 
so, in order to make universality vivid.

All I need is to report the striking, indeed amazing, fact that the two members  of each 
pair ((i) and (ii), (i') and (ii')) {\em obey the same law}. That is: (i) and (ii) obey the  
same power-law, namely with $p \approx 0.35$. In fact this $p$ is written as $\beta$. So we 
have
\be\label{cretbeta}
\rho_{\rm{liquid}} - \rho_{\rm{gas}} \sim |T - T_c|^{\beta} \;\;\; {\mbox{and}} \;\;\; m 
\sim |T - T_c|^{\beta}; \;\;\; {\mbox{with}} \;\;\; \beta \approx 0.35.
\ee
Furthermore, this power-law, with almost the same value of $\beta$, describes corresponding 
phase transitions in many other systems. Here is an example like (i): as one reduces  the 
temperature of a sample of liquid helium  below 2 K (two degrees above absolute zero, i.e. 
-271 degrees Celsius) another phase of helium (a superfluid phase: so-called because it 
flows without viscosity) condenses out, rather like liquid water condensing out of steam. 
This transition has a power law like eq. \ref{cretbeta}  with $\beta \approx 0.34$. Another 
example like (ii) is the magnetization of nickel: here, $\beta \approx 0.36$.

Similarly, (i') and (ii') obey the same power-law, with nearly equal values of their 
exponent $p$: namely $p \sim - 1.2$. By convention, this $p$ is written as $- \gamma$, so 
that $\gamma$ is positive. So we have
\be\label{cretgamma}
\kappa_T  \sim |T - T_c|^{- \gamma} \;\;\; {\mbox{and}} \;\;\; \chi_T  \sim |T - T_c|^{- 
\gamma}; \;\;\; {\mbox{with}} \;\;\; \gamma \approx 1.2.
\ee
Again, this power-law, with almost the same value of $\gamma$, describes corresponding phase 
transitions in many other systems. Here is an example like (i'): a mixture of two organic 
fluids (viz. trimethylpentane and nitroethane) has a critical point rather like that of 
water-and-steam: the compressibility $\kappa_T$ obeys a power law like eq. \ref{cretgamma},  
with $\gamma \approx 1.24$. And for helium, i.e. our previous example like (i): the 
compressibility has $\gamma \approx 1.33$. Another example like (ii') is the magnetic 
susceptibility of nickel: here, $\gamma \approx 1.33$.

$\beta$ and $\gamma$ are called {\em critical exponents}. By convention, critical exponents 
are defined to be  positive; so that if the quantity concerned diverges at the critical 
point, the exponent/index in the power-law is minus the critical exponent. There are several 
others. All of them encode some power-law behaviour of quantities at or near a critical 
point: and each of them takes very nearly the same value for critical points occurring in 
materials and processes that are microscopically very disparate. So again, we see a striking 
universality of systems' quantitative (albeit arcane!) behaviour.

I need not go into details, except to describe how two critical exponents, written $\eta$ 
and $\nu$, encode the power-law behaviour near the critical point---not of  
straightforwardly macroscopic quantities like the density, compressibility etc. so far 
mentioned---but of {\em measures of the microscopic correlations}. Sketching this will bring 
us back to Section \ref{crpt}'s central idea of scale-invariance.

One of course expects correlations between the states at different locations in a material 
to decrease, as one considers more widely separated locations. Indeed, this is so: and at or 
near critical points, the decrease  is often given by a power of the separation. Thus 
suppose we define as our measure of correlation, the expectation (probabilistic average) of 
the product of local variables (often called `order parameters') at the two locations, 
separated by distance $r$. This is often written $G(r)$.\footnote{It is usually defined in 
terms of local variables $\phi$ by: (i) assuming spatial homogeneity so that any two 
locations a distance $r$ apart yield the same expectation, and one may as well choose the 
origin $\bf 0$ and some location $\bf r$ which is $r$ units away, so that one would  write 
$G(r) = \langle \phi({\bf 0}) \cdot \phi({\bf r}) \rangle$, where $\langle  \rangle$ 
indicates expectation; and (ii) subtracting away any background constant correlation in 
$\phi$, the better to study fluctuations, so that one writes $G(r) := \langle \phi({\bf 0}) 
\cdot \phi({\bf r}) \rangle - | \langle \phi \rangle |^2$.}  At the critical temperature 
$T_c$, $G(r)$ obeys a power law, for $r$ large compared with inter-molecular distances.  
Namely:
\be\label{creteta}
G(r) \sim  1/r^{d - 2 + \eta}
\ee
where (i) $d$ is the spatial dimension of the material (usually 3; but e.g. 2 for a 
monoatomic or monomolecular layer, or a theoretical model of such), and (ii) $\eta$ is 
typically between 0 and 0.1. (In my examples, it is: (a) for the mixture of two organic 
liquids and for helium, 0.02; (b) for iron, 0.07; (c) for nickel, 0.04.)

Near $T_c$, $G(r)$ obeys, not a simple power law in $r$, but an exponential decrease:
\be\label{xidefine}
G(r) \sim  \exp (-r/\xi) ;
\ee
which defines the length $r = \xi$ over which $G(r)$  decreases by about 66 $\%$ ($e \approx 
2.7$ so that $\exp (- \xi/\xi) \equiv e^{-1} \approx 1/3$). This means that $\phi$ 
fluctuates in correlated blocks of various sizes up to a length $\xi$, but blocks with a 
side-length much bigger than $\xi$ are very rare.

It is here that we see again  the {\em scale-invariance} of the critical point---and so its 
being an infra-red fixed point of the renormalization group flow. For as $T_c$ is approached 
from above or from below, $\xi$ grows without limit. Indeed, its behaviour is described by 
(yet another!) power law: with a negative exponent so as to capture the divergence, as with 
the $- \gamma$ for (i') and (ii') above. That is: near $T_c$ (i.e. $| T - T_c |/T_c << 1$), 
we have
\be\label{cretnu}
\xi \sim | T - T_c |^{- \nu}
\ee
where $\nu$ is typically about 2/3. Again, the approximate value 2/3 covers a whole class of 
phenomena: for our four examples of the critical points for the two organic liquids, helium, 
iron and nickel, the range of $\nu$ is from 0.62 (the liquids) to 0.69 (iron).\footnote{This 
discussion, especially the data cited, is based on Binney et al. (1992, pp. 7-8, 18-20, 
22).}\\

Finally, a note of caution.  My praise of the modern approach to renormalization, with its 
successful calculation of critical exponents (and, needless to say: so much else!), might 
give the impression that only with the renormalization group did physics cotton on to the 
broad idea of coarse-graining a microscopic description to give a macroscopic one, or of 
iterating a scheme for coarse-graining, so as to understand scale-invariant phenomena, and 
in particular to calculate critical exponents.

That is {\em not so}. Before the renormalization group analyses were developed, various 
approaches to understanding these phenomena had considerable, albeit partial, success. That 
is of course hardly surprising, since the ideas of coarse-graining, and iterating a scheme 
for coarse-graining, are so natural. But it is worth emphasizing: not just so as to honour 
the previous work, but  also to avoid philosophers getting the false impression that only 
with the renormalization group did physics pick up on the idea of multiple realizability. In 
short: the renormalization group is of course a scientific triumph, and a stimulus to 
philosophical relection---but that does {\em not} imply that it harbours entirely novel 
morals for philosophy. Section \ref{Nag} will expand on this irenic theme.\footnote{Two 
broad previous approaches are mean field theory and Landau theory. For a glimpse of what 
these are and their predictive limitations, in the context of condensed matter, cf. e.g. 
Kadanoff (2009, Sections 2,3; 2013, Sections 1.2.4-4, pp. 147-164) and Binney et al. (1992, 
pp. 22, 176-177, 183-184). In particular, mean field theory implies values for the critical 
exponents I have discussed, as follows: $\beta = 0.5, \gamma = 1, \eta = 0, \nu = 0.5$. As 
to the false impression that within physics only the renormalization group describes 
multiple realizability: I think this impression is fostered by some philosophers' praise of 
the renormalization group; e.g. Batterman (2002, pp. 37-44), Morrison (2012, p. 156, p.160 
(both paragraph 2)).}

\subsubsection{Short distances: a natural lower limit}\label{latt}
This Section has stressed the analogy between the renormalization group framework for 
quantum field theory and for condensed matter. But I should end by noting two main 
differences; the second will be more important. (Of course, there are also many differences 
of detail.)

First: I have spoken impartially of the renormalization group flow going (i) towards long 
distances, the infra-red (Sections \ref{dwindle} and \ref{crpt}), and (ii) towards short 
distances, the ultra-violet (Section \ref{beta}). But broadly speaking, a flow towards the 
infra-red is often defined by (iterated) coarse-graining. And since coarse-graining involves 
a loss of microscopic information, such a flow is in general many-one, i.e. irreversible: 
this underpinned Section \ref{crpt}'s idea of two disparate microscopic Hamiltonians flowing 
to the same infra-red fixed point. So there is no well-defined inverse flow towards the 
ultra-violet. (And in the jargon of group theory, one should thus say `renormalization 
semi-group', not `renormalization group'; but the latter name has stuck.) This point is not 
meant to suggest there is a problem, or even a puzzle, about such irreversible flows.  The 
exact definition of the flow is up to the theorist, and of course varies from theory to 
theory, and according to the theorist's purposes. Thus the flow towards the infra-red is not 
always defined by coarse-graining (loss of information);  so in some cases it is indeed 
reversible, and there is a well-defined flow to the ultra-violet.\footnote{Cf. e.g. Kadanoff 
(2013, Section 8.5, p.181). I say there is `not even a puzzle'; but it can be 
confusing---witness the exchanges between Cao and Fisher which pepper Cao (1999a).}

The second difference is about the ultra-violet regime. Namely: quantum field theory lacks, 
but condensed matter has, a natural lower limit to the lengths $d$ and $L$. In more detail: 
quantum field theories, being {\em field} theories, assume a spacetime continuum. Although I 
pointed out that we can make sense of such theories without spacetime being a continuum, 
namely as describing physics at low enough energies, large enough distances (Sections 
\ref{notcontm} and \ref{eft?}), these theories are undoubtedly conceived and presented as 
involving a continuum. After all, that was the rationale for the core task, on the 
traditional approach to renormalization in Sections \ref{cut} and \ref{dto0}: to assign bare 
coupling constants at each value of the cut-off $d$, so as to recover the physical coupling 
constants (cf. eq. \ref{limg0}). Similarly for $L$ and $\mu$. In a theory assuming a 
spacetime continuum, there is no {\em a priori} obstacle to arbitrarily short-distance, 
high-energy phenomena---$L$ can be arbitrarily small, $\mu$ arbitrarily high.

But in condensed matter, the inter-atomic or inter-molecular distance gives a natural lower 
limit to the length-scales at which we trust our analyses. Thus recall that in Section 
\ref{crpt}, I said the critical point involves scale-invariance, `until we reach molecular 
dimensions, where, of course, the alternation between the phases breaks down'. More 
generally: we of course expect our definition of the renormalization group flow, by which 
one ``vector'' of physical coupling constants (or: one Hamiltonian or Lagrangian) replaces 
another as the energy scale increases (distance scale decreases) to break down at molecular 
or atomic scales. (Or of course at larger length-scales, if some new physics ``kicks in'' 
there, which our definition of the flow has disregarded: an example would be spatially  
periodic inhomogeneities with, say, a mesoscopic period.) Similarly of course, for 
theoretical models rather than real samples. The classic example is modelling a crystal as a 
spatially periodic lattice of sites representing atoms or nuclei. In such a model, the 
lattice spacing provides the natural lower limit. To sum up: at this limit, the 
renormalization group transformation---the step along the flow towards the ultra-violet  
that ``zooms in'' by an increment---cannot be iterated.

Mention of lattice models in condensed matter prompts a final remark, to complete my 
description  of quantum field theory. Namely, lattice models are also crucial there: but 
with a lattice of sites throughout a region of  spacetime, rather than space. We can easily 
state two broad reasons why.

(1): Since quantum field theories are complicated, one often cannot solve one's problem 
analytically (even with perturbation theory, power series etc.) and so needs to solve it 
numerically, i.e. on a computer: which involves discretizing the field variables, i.e. 
assigning them to sites of a lattice rather than the continuum of spacetime points.

(2): I enthused about QCD's asymptotic freedom: that the theory becomes very simple, indeed 
free, at arbitrarily high energies. But consider the other side of this coin: as the energy 
is lowered, the interaction becomes stronger. Of course it was realized already in the 1930s 
that the interaction holds the nucleus together, overcoming the electrostatic repulsion 
among the protons. Hence its name, `strong', given long before QCD and asymptotic freedom 
were found. This strength means a large physical coupling constant, and so difficulties for 
a perturbative expansion (cf. Section \ref{vpptbn}'s discussion of $\varepsilon$ and 
$\alpha$): difficulties that led in the 1950s and 1960s to scepticism that any perturbative 
quantum field theory could describe the interaction.  Again there is jargon: the increasing 
strength at larger distances is called  {\em infra-red slavery}, and the ever-tighter 
binding of quarks as they try to separate (which prevents us seeing a lone quark) is called 
{\em confinement}; (e.g. Wilczek 2005, Sections 1.1, 3.1). But jargon aside, the predicament 
that the theory is complicated and very hard to calculate in prompts lattice models and the 
hunt for numerical solutions.

\section{Nagelian reflections}\label{Nag}
So much by way of reviewing renormalization, in both the traditional and modern approaches. 
In this final Section, I turn briefly to philosophy. I will confine myself to urging a moral 
which honours Ernest Nagel. Namely: the material in Sections \ref{tradnut} and \ref{modnut} 
fits well with Nagel's account of reduction; and indeed, Hempel's deductive-nomological 
account of explanation. In particular, the idea of universality (Sections \ref{flow} and 
\ref{cret}) is a case of the philosophical idea of multiple realizability, and does not 
cause problems for Nagelian reduction.

  That is good news for me, since in previous work I 
joined others in endorsing Nagel's account, and in arguing that multiple realizability is no 
threat to Nagelian reductions  (2011, Sections 3 and 4; and for renormalization, 2014). I will not repeat any of the 
details of those arguments here.\footnote{\label{fansNagel}{Not that Nagel's doctrines about explanation, 
reduction and inter-theoretic relations need {\em my} endorsement: especially since, 
happily, his stock-price is rising. For example, one recent defence of Nagel's account, and 
similar accounts such as Schaffner's (1967, 1976), is Dizadji-Bahmani, Frigg and Hartmann 
(2010); and Schaffner has recently given a masterly review of this literature, and defence 
of his own account (2013). As to multiple realizability not threatening Nagelian reduction, 
I endorse  Sober (1999).}} I will just sketch Nagel's account (Section \ref{manna0}) and 
describe how the material I have reviewed illustrates it (Section \ref{manna1}). Finally in 
Section \ref{manna2} I will urge that universality is just multiple realizability.

But before setting out, I note as an appetizer for another occasion, two philosophical  
topics prompted by the material in Sections \ref{tradnut} and \ref{modnut}.

(1): {\em Real infinities?}: Should we acquiesce in the traditional approach's acceptance of 
infinite bare coupling constants (Section \ref{dto0})? I noted there that most physicists 
find such physically real, albeit unmeasurable, infinities intellectually uncomfortable; and 
for all the achievements of the modern approach (especially Sections \ref{dwindle}, 
\ref{cret}), this approach does {\em not} expunge these infinities. So what should a 
philosopher say about such infinities: are they acceptable?\footnote{In some of his work, Batterman has advocated physically real infinities (e.g. 2005, 
pp. 235-236). But beware: this is a different topic. For it rests on different considerations than 
infinite bare coupling constants in renormalization theory: viz. (i) singular limits, and 
(ii) the need to take a limit, so as to describe or explain a distinctive phenomenon, e.g. 
the thermodynamic limit so as to get a phase transition. For my own assessment of his 
arguments about (i) and (ii), cf. my (2011a, Section 3; 2014, Section 5.1).}

(2): {\em Instrumentalism?}: The dwindling of non-renormalizable terms in the infra-red 
prompted the effective field theory programme, with its vision of a succession of theories, 
each accurately describing one energy range, but inaccurate beyond it---suggesting an 
instrumentalist attitude (Section \ref{eft?}). But of course, this dwindling (and allied 
technical considerations) do not make such an attitude compulsory, especially if we think 
there are good prospects for getting an empirically adequate and rigorously defined 
interacting quantum field theory: prospects that are certainly enhanced by QCD's asymptotic 
freedom (cf. the end of Section \ref{beta}). So what attitude should we adopt? The answer is bound to be controversial, since it is an 
example of the broad, and maybe unending, philosophical debate between varieties of 
instrumentalism and of realism.\footnote{For more discussion, with an emphasis on 
renormalization and the idea of ``reductionism'', cf.: (i) Cao (1993), Schweber (1993), Cao and Schweber (1993),  Cao (1997, 
Section 11.4 pp. 339-352); (ii) the brief contrasting discussions by 
Weinberg (1999, pp. 246-250) and Redhead (1999, pp. 38-40); (iii) responses to Cao and Schweber by Hartmann (2001) and Castellani (2002); and more recently, (iv) Bain (2013, 2013a) and my (2014, Section 5.3) .}


\subsection{Nagel endorsed}\label{manna0}
The traditional philosophical jargon is that one theory $T_1$ {\em reduces} to, or {\em is 
reducible to}, another $T_2$ if, roughly speaking, $T_1$ is a {\em part} of $T_2$. To 
explicate this,\footnote{Of course, we can and should admit that `reduction', used as a 
relation between theories, is vague (my 2011, Section 1.1.2). But this is undoubtedly the 
core idea.} I first introduce mnemonic subscripts, to avoid confusion between $T_1$ and 
$T_2$.

For the reducing theory, I will use $b$: `b' is for bottom/basic/best; `bottom' and `basic' 
connoting microscopic and fundamental, and `best' connoting a successor theory. And  for the 
reduced theory, I will use $t$: `t' is for top/tangible/tainted; `top' and `tangible' 
connoting macroscopic and observable, and `tainted' connoting a predecessor theory. Thus I 
will  say that a
theory $T_b \equiv T_{\rm{bottom/basic/best}}$ reduces $T_t \equiv 
T_{\rm{top/tangible/tainted}}$; and that $T_t$ reduces to, or is reducible to, $T_b$.

Nagel's main idea is to take theories as linguistic entities, viz. sets of sentences closed under 
deducibility, and hold  that $T_t$ is reduced to $T_b$ by: \\
\indent (1) {\em Deducibility}: being shown to be  logically deducible from $T_b$, usually 
together with some judiciously chosen definitions; and  \\
\indent (2) {\em Constraints}: the deduction satisfying some informal constraints,  e.g. 
about the derivation counting an explanation of its conclusion.

Before I give some details about (1) and (2), in Sections \ref{defext} and 
\ref{philcritdefext} respectively, my first and most important comment is that taking theories as sets of sentences does {\em not} imply that the language of the theories, or the notion of deducibility, be formalized (let alone  
first-order). Nagel and his supporters (including Schaffner et al. listed in footnote \ref{fansNagel}) of course  
know that scientific theories are not formalized, and are hardly likely to be, even in  
mathematized subjects like physics. But that does not prevent one undertaking to assess  
whether there are relations of reduction between theories (in Nagel's sense; or indeed, in  
another one). The informality merely makes one's specification of the theories, definitions and deducibility relation,   
somewhat vague; and so one's ensuing  
assessments are correspondingly tentative.\footnote{I agree that  the tenor of Nagel's writings---and of his times---also suggests a narrower, and I admit 
implausible, account requiring that theories, definitions and deductions, be formalized.  (Thanks to Bob Batterman and Jim Weatherall for 
emphasizing this.) I say a bit more about this, and the rival ``semantic view'' of theories in my (2011, Sections 3, 4; 2014, Section 1.2). I should also note that, on the other hand, some writers have for various reasons criticized the notion of a scientific theory, even if theories are not taken as linguistic entities; and have even invoked quantum field theories in their support. I reply to this line of thought, in the companion paper (2014, Section 4.1).}

\subsubsection{Deducibility}\label{defext}
 The syntactic conception of theories immediately gives a notion of $T_t$ being a {\em part} 
of $T_b$, viz.\ when $T_t$'s sentences, i.e. theorems, are a subset of $T_b$'s.

 In logic, with its formalized languages, this is called $T_t$
being a {\em sub-theory} of $T_b$. But the notion of course applies to unformalized 
languages. One just has to recognize that which sentences are in a given theory is almost 
always vague, even when the theory's name is very specific (e.g. `equilibrium classical 
statistical mechanics of $N$ point-particles in a cube with side-length $l$', rather than 
`classical statistical mechanics'). And  to avoid merely verbal disputes, one must adopt a 
swings-and-roundabout attitude about whether a given $T_t$ is part of a given $T_b$. For it 
all depends on how one makes the two theories, i.e. their sets of sentences, precise; (cf. 
my 2011, Section 3.1.2; and Schaffner (2013, Section II p. 7, Section VI, p. 29)).

So although scientific theories are of course unformalized, I will continue to use logical 
jargon, saying e.g. that a theory has a set of sentences, a language has a set of predicates 
etc. I will also set aside the questions of what a theory's underlying logic is (first-order 
vs. second-order etc.) and what its mathematical apparatus is (containing set theory, 
calculus etc.), since for unformalized theories these questions also are vague. So I just 
refer non-commitally to `deduction'.

However, one still needs to avoid confusion arising from the same predicate, or other
non-logical symbol, occurring in both theories, but with different intended interpretations. 
This can be addressed by taking the theories to have disjoint
non-logical vocabularies, and then defining $T_t$ to be a
{\em definitional extension\/} of $T_b$, iff one can add to
$T_b$ a set $D$ of definitions, one for each of $T_t$'s non-logical symbols,
in such a way that $T_t$ becomes a sub-theory of the augmented theory $T_b \cup D$.
That is: In the augmented theory, we
can deduce every theorem of $T_t$.

Note that here again, I mean the word `definition'  as it is used in logic: roughly speaking, a definition 
states that a given ``new''  non-logical symbol e.g. a predicate, the {\em definiens}, is co-extensive with some (maybe very long) open sentence, the {\em definiendum}. There is 
no formal requirement that a definition be faithful to some previous meaning (or some other 
properties) of the definiens: I shall return to this in Section \ref{philcritdefext}.A.

This is the core formal part of Nagelian reduction: (1) {\em Deducibility}, in the above 
notation. The definitions are usually called `bridge laws' (or `bridge principles' or 
`correspondence rules').\footnote{The standard references are Nagel (1961, pp. 351-363; and 
1979); cf. also Hempel (1966, Chapter 8). Discussions of bridge laws are at Nagel (1961, pp. 
354-358; 1979, 366-368) and Hempel (1966, pp. 72-75, 102-105). My notation {\em 
Deducibility} thus covers the conditions Nagel calls `derivability' (about logical 
consequence) and `connectability' (about definitions); cf. the discussion in Schaffner 
(2013, Section I). In Section \ref{philcritdefext} I will mention revisions by authors such 
as Schaffner.}

\subsubsection{Informal constraints on reduction}\label{philcritdefext}

\paragraph{4.1.2.A: Deducibility is too weak}\label{412.A} Various philosophers have said 
that even if $T_t$ is
a definitional extension of $T_b$, there can be non-formal aspects of $T_t$ that are
not encompassed by (are not part of) the corresponding
aspects of $T_b$; and that, at least in some cases, these aspects
seem essential to $T_t$'s functioning as a scientific theory. Nagel agrees with the general point that deducibility is too weak, and (1961, pp. 358-363) 
adds  some informal
conditions, mainly motivated by the idea that the reducing
theory $T_b$ should explain the reduced theory $T_t$; and following
Hempel, he conceives explanation in deductive-nomological
terms. Thus he says, in effect, that $T_b$ reduces $T_t$
iff: \\
\indent (i): $T_t$ is a definitional extension of $T_b$; and \\
\indent (ii): in each of the definitions of $T_t$'s terms, the {\em definiens} in the 
language of $T_b$ must play
a role in $T_b$; so it cannot be, for example, a
heterogeneous disjunction.

Nagel's proposal (ii) has got some good press; e.g. Sklar (1967, pp. 119-123), Nickles 
(1973, pp. 190-194). But some philosophers neglect this proposal and-or object that it does not secure what is needed. The most common objection appeals to {\em multiple realizability}. The leading idea is that the {\em 
definiens} of a multiply realizable property shows it to be too ``disjunctive'' to be 
suitable for scientific explanation, or to enter into laws. And {\em pace} Nagel's proposal 
(ii) above, some philosophers think that multiple realizability prompts a non-Nagelian
account of reduction.

These lines of thought are most familiar in the philosophy of mind, with the {\em multiple 
realizability argument} of Putnam (1975) and Fodor (1974). Kim's work provides an example 
which specifically targets Nagelian reduction. He
calls it `philosophically empty' and `irrelevant', because Nagelian bridge laws are `brute 
unexplained primitives'; (1999, p. 134; 2006, p. 552; and similarly in other works e.g. 
(2005, pp. 99-100): cf. also e.g. Causey (1972, p. 412-421)). Kim's own account, called a 
`functional model of reduction', takes reduction to include:\\
\indent (a) functional definitions of the higher-level properties $P$ etc., i.e. definitions that quantify over properties so as to characterise $P$ by its pattern of relations (usually causal and-or nomic) to various other properties; and \\
\indent  (b) a lower-level description of the (variety-specific) realizers of $P$ etc., and of how they fulfill the functional roles spelt 
out in (a).  

But as announced, I reject these lines of thought. I believe multiple realizability, and functional definitions, give no 
argument against definitional extension; nor even against stronger notions of reduction like 
Nagel's, that add further constraints additional to deducibility, e.g. about explanation. 
That is: I believe that such constraints are entirely compatible with multiple 
realizability. I maintain (2011, Sections 3.1.1 and 4.1.1) that this was shown 
persuasively by Sober (1999).\footnote{Sober mostly targets Putnam and Fodor; for critiques 
of Kim, cf. Marras (2002, pp. 235-237, 240-247), Needham (2009, pp. 104-108)}. I will not 
repeat the rebuttal. Let me just sum up: there is no tension between Nagelian reduction and 
the fact of multiple realizability.

\paragraph{4.1.2.B: Deducibility is too strong}\label{412B} The other traditional criticism 
of Nagelian reduction goes in the ``opposite direction''. Instead of saying that Deducibility is too
weak, it says that Deducibility is too strong. The idea is 
that in many cases where $T_b$ reduces $T_t$, $T_b$ {\em corrects}, rather than implies, 
$T_t$.\footnote{This 
objection is made by e.g. Kemeny and Oppenheim (1956), and Feyerabend (1962).  One standard example is Newtonian gravitation theory ($T_b$) and Galileo's
law of free fall ($T_t$). This $T_t$ says that bodies near the earth fall with constant 
acceleration. This $T_b$ says that as they fall, their acceleration increases, albeit by a 
tiny amount. But surely $T_b$ reduces $T_t$. And similarly in many famous and familiar 
examples of reduction in physics: wave optics corrects geometric optics, relativistic 
mechanics corrects Newtonian mechanics etc.

These examples put limiting relations between theories centre-stage. And quite apart from 
assessing Nagel, such relations have long been a major focus for philosophers of the 
physical sciences: as also in recent years, cf. e.g. Batterman (2002, 2013), Norton (2012). 
My own views about these relations are in (2011a): which takes as one of its examples, phase 
transitions, especially Section \ref{crpt}'s topic of critical points; (cf. also Bouatta and 
Butterfield (2011, 2012).}

But as an objection to Nagel, this is an {\em ignoratio elenchi}. For Nagel himself made this point. He said that a case
in which $T_t$'s laws are a close approximation to
what strictly follows from $T_b$ {\em should} count as reduction. He called this {\em 
approximative reduction} (1979, pp. 361-363, 371-373). (Cf. also  Hempel (1965, p. 344-346; 
1966, pp. 75-77). In a similar vein, Schaffner (1967, p. 144; 1976, p. 618) requires a 
strong analogy between $T_t$ and its corrected version, i.e. what strictly follows from 
$T_b$). 

Against this, critics have complained that it is too programmatic, since it gives no 
general account of when an approximation or analogy is close enough. But in Nagel's and 
Schaffner's defence, I would say  that (a) we
should not expect, and (b) we do not need, any such general account.\footnote{I am not alone 
in this defence: cf. Nickles (1973, p. 189, 195), Schaffner (2013, Section III) and 
Dizadji-Bahmani, Frigg and Hartmann (2010: Section 3.1, p. 409f.). In a similar vein, the 
latter argue (their Section 5) that Nagelian reduction does not need to settle once for all 
whether bridge laws state ``mere'' correlations, law-like correlations or 
property-identities: (which are Nagel (1961)'s three options, at pp. 354-357). I entirely 
agree, {\em pace} authors like Kim and Causey cited in Section 
\ref{philcritdefext}.A.} What {\em matters}, both scientifically and conceptually, is that 
in a given case we can deduce that $T_t$'s proposition (whether a description of particular 
fact, or a general theorem/law) is approximately true; and that we can quantify how good the 
approximation is.

\subsection{Renormalizability deduced at low energies as a family of Nagelian 
reductions}\label{manna1}
I now conclude by discussing how renormalization illustrates Nagelian reduction. I will 
confine myself to the topic announced in Section \ref{intro}: how the modern approach, with 
its explanation that our good fortune in having renormalizable theories is generic (Section 
\ref{dwindle}), accords with Nagelian reduction.

As I said in Section \ref{intro}, I think philosophers should take note, not just of this 
specific achievement, but of the general idea of a {\em space of theories}. This fosters a 
novel and more ambitious kind of explanatory project than the familiar ones of explaining an 
individual event,  or  a single law, or (as in Section \ref{manna0}) a theory as a part of, 
or a good approximation to, another. Namely: to explain a feature of a whole class of 
theories in a unified way in terms of the structure of the space of theories.

But I will not develop this general theme, but instead concentrate on my main claim, as 
follows:
\begin{quote}
The deduction from a given theory $T_b$ that describes (perhaps using non-renormalizable 
terms) high-energy physics, of a renormalizable theory $T_t$ describing low energy physics, 
is a Nagelian reduction. Besides: for different pairs of theories $T_b$ and $T_t$, varying 
across the space of quantum field theories, the reductive relation is  similar, thanks to a 
shared definition of the renormalization group flow, i.e. of the renormalization scheme.
\end{quote}
I will fill this out with five short remarks, rehearsing previous material. As regards the 
philosophical assessment of Nagel's account of reduction, the most important of these 
remarks are (3), about approximative reduction, and (4), about unity among a family of 
reductions.

(1): I have specified a theory by a Hamiltonian or Lagrangian. Recall the vector of physical 
coupling constants $g_1(\mu),..., g_N(\mu)$ which I first introduced in Section 
\ref{physrat}, and took as a point in a space of theories at the start of Section 
\ref{modnut}. 

(2): A renormalization scheme that defines a flow towards lower energies (a scheme  for 
coarse-graining so as to eliminate higher-energy degrees of freedom) amounts to the set of 
definitions $D$, in logicians' broad sense (Section \ref{defext}), needed to make the deduction of $T_t$ from $T_b$ go 
through.

(3): Since at low energies any non-renormalizable terms in $T_b$ still make non-zero, albeit 
dwindling, contributions to the theory's predictions (probabilities for various processes), 
we have here an approximative reduction (Section \ref{philcritdefext}.B); though the 
approximation gets better and better as the energy decreases.

(4): A given renormalization scheme (definition of the flow) works to show that many 
theories $T_b$ lead to a renormalizable low-energy theory $T_t$ . This unity is striking. 
Hence this Section's title's use of the word `family': since `family' connotes resemblance, 
which `set' does not.

(5): Agreed: no single renormalization scheme works to prove that all possible theories have 
dwindling non-renormalizable contributions in the infra-red. And as I have often admitted: 
the proofs concerned are often not mathematically rigorous. But the various renormalization 
schemes that have been devised do not contradict one another, and in fact mesh in various 
ways. So it is fair to say that a generic quantum field theory has dwindling 
non-renormalizable contributions in the infra-red. As I said in Section \ref{dwindle}: a 
satisfying explanation of our good fortune.\footnote{That a satisfying explanation should 
show the {\em explanandum} to be generic has of course been a theme in cosmological 
speculations about explaining initial conditions: not just nowadays, but over the centuries; 
cf. McMullin (1993).}

So to sum up: the modern approach to renormalization gives a stunning case of explaining 
something that is otherwise---i.e. on the traditional approach---a coincidence. The 
coincidence is, in short, that the theory in question, e.g. quantum electrodynamics, has a 
feature, viz. renormalizability, that seems crucial to it ``making sense''. This feature 
also holds of other theories;  indeed, it holds of the whole standard model of particle 
physics, which combines quantum electrodynamics with quantum theories of the weak and strong 
forces. Thus the coincidence is so grand and important that it seems like manna from heaven; 
or more prosaically, it seems that renormalizability is in some sense an {\em a priori} 
selection-principle for quantum theories. But adopting the modern approach, we can deduce 
that what seemed manna from heaven is in a sense to be expected.

\subsection{Universality is multiple realizability}\label{manna2}
My final point is a brief ancillary one. In Section \ref{flow}, I introduced `universality' 
as jargon for the idea that dissimilar theories might have similar infra-red behaviour: in 
particular, the same infra-red fixed point. In Section \ref{cret}, we saw vivid examples of 
this. The dissimilar theories were of utterly different quantities in utterly different 
systems: e.g. a density difference in a mixture of water and steam, or in a mixture of two 
phases of liquid helium; or the magnetization of a piece of iron or nickel. And the similar 
behaviour was quantitatively exact, albeit arcane: the same values of critical exponents in 
power laws describing near-critical behaviour. In the examples just mentioned, it was the 
critical exponent $\beta$ in eq. \ref{cretbeta}. Corresponding remarks apply to the 
exponents  $\gamma$, $\eta$ and $\nu$, in eq. \ref{cretgamma}, \ref{creteta} and 
\ref{cretnu} respectively.

These examples obviously illustrate multiple realizability in the sense introduced in 
Section \ref{philcritdefext}.A. To take just my first example: having a value of $\beta$ equal to 
about 0.35 as in eq. \ref{cretbeta} is a multiply realizable property. A water-steam mixture 
has it; so does a mixture of two phases of liquid helium. Similarly for Section 
\ref{philcritdefext}.A's idea of a functional definition of a property, viz. specifying a pattern of relations to other properties: clearly, having $\beta \approx 0.35$ is a functional 
property. 

But recall that I join Sober (1999) in seeing no tension 
between Nagelian reduction and multiple realizability, with the consequent need for 
functional definitions. So I also see no trouble for Nagelian reduction in universality: e.g. in critical exponents, whether calculated using the renormalization group or using older approaches like mean field theory (cf. footnote 19). In particular, I see no trouble for Section \ref{manna1}'s main 
claim that to coarse-grain a class of theories towards their common infra-red fixed 
point---zooming out our descriptions in each theory, towards longer distances---is to give a 
unified family of Nagelian reductions. \\ \\

{\em Acknowledgements}:--- JB thanks: Sebastien Rivat, Nic Teh, and seminar audiences in Cambridge for discussion and comments on early versions (2012); Jonathan Bain, Bob Batterman, Tian Yu Cao, Elena Castellani, and Jim Weatherall for generous comments on a later draft (2013); and audiences at Columbia University, New York and Munich (2013).  Thanks also to the editors and a very perceptive referee. While I have incorporated most of these comments, I regret that I could not act on them all. I thank the {\em Journal of Philosophy} for allowing some overlap (in Sections 3.1.2---3.1.4 and 4.2) with the companion paper (2014). This work was supported in part by a grant from the Templeton World Charity Foundation, whose support is gratefully acknowledged.

\section{References}
Aitchison, I. (1985), `Nothing's plenty: the vacuum in quantum field theory', {\em 
Contemporary Physics} {\bf 26}, p. 333-391.

Applequist, T. and Carazzone, J. (1975), `Infra-red singularities and massive fields', {\em 
Physical  Review} {\bf D11}, pp. 2856-2861.

Baez, J. (2006), `Renormalizability': (14 Nov 2006):\\
 http://math.ucr.edu/home/baez/renormalizability.html

Baez, J. (2009), `Renormalization made easy': (5 Dec 2009): \\ 
http://math.ucr.edu/home/baez/renormalization.html

Bain, J. (2013), `Effective Field Theories', in Batterman, R. (ed.) {\em The
Oxford Handbook of Philosophy of Physics}, Oxford University Press, pp.
224-254.

Bain, J. (2013a), `Emergence in Effective Field Theories', {\em European Journal
for Philosophy of Science}, {\bf 3}, pp. 257-273. (DOI 10.1007/s13194-013-0067-0)

Batterman, R. (2002), {\em The Devil in the Details}, Oxford University Press.

Batterman, R. (2005), `Critical phenomena and breaking drops: infinite idealizations
in physics', {\em Studies in History and Philosophy of Modern Physics} {\bf 36B}, pp. 
225-244.

Batterman, R. (2010), `Reduction and Renormalization',  in A. Huttemann and G. Ernst, eds.  
{\em Time, Chance, and Reduction: Philosophical Aspects of Statistical Mechanics}, Cambridge  
University Press, pp. 159--179.

Batterman, R. (2013), `The tyranny of scales', in {\em The Oxford Handbook of the Philosophy 
of Physics}, ed. R. Batterman, Oxford University Press, pp. 255-286.

Binney, J., Dowrick, N, Fisher, A. and Newman, M. (1992), {\em The Theory of Critical 
Phenomena: an introduction to the renormalization group}, Oxford University Press.

Bouatta, N. and Butterfield, J. (2011), `Emergence and Reduction Combined in Phase 
Transitions', in J. Kouneiher, C. Barbachoux and D.Vey (eds.), {\em Proceedings  of 
Frontiers of Fundamental Physics 11} (American Institute of Physics); pp.?? . Available at: 
http:// philsci-archive.pitt.edu/8554/ and at: http://arxiv.org/abs/1104.1371

Bouatta, N. and Butterfield, J. (2012), `On emergence in gauge theories at the 't Hooft 
limit', forthcoming in  {\em European Journal for Philosophy of Science},  http://philsci-archive.pitt.edu/9288/

Brown, L. ed. (1993), {\em Renormalization: from Lorentz to Landau (and Beyond)}, Springer.

Butterfield, J. (2004), `Some Aspects of Modality in Analytical Mechanics', in P. 
Weingartner and M. Stoeltzner (eds), {\em Formale Teleologie und Kausalitat in der Physik}, 
Mentis 2004; pp. 160-198;\\
 physics/0210081; and http://philsci-archive.pitt.edu/archive/00001192/

Butterfield, J. (2011), `Emergence, Reduction and Supervenience: a Varied Landscape', {\em 
Foundations of Physics}, {\bf 41}, pp. 920-960. At Springerlink: doi:10.1007/s10701-011-9549-0; 
http://arxiv.org/abs/1106.0704: and at: http://philsci-archive.pitt.edu/5549/

Butterfield, J. (2011a), `Less is Different: Emergence and Reduction Reconciled', in {\em 
Foundations of Physics} {\bf 41}, pp. 1065-1135. At: Springerlink (DOI 
10.1007/s10701-010-9516-1); http://arxiv.org/abs/1106.0702;
and at: http://philsci-archive.pitt.edu/8355/

Butterfield, J. (2014), `Reduction, emergence and renormalization',  {\em The 
Journal of Philosophy} {\bf 111}, pp. 5-49. 

Cao, T.Y. (1993), `New philosophy of renormalization: from the renormalization group to 
effective field theories', in Brown ed. (1993), pp. 87-133.

Cao, T.Y. (1997), {\em Conceptual Developments of Twentieth Century Field Theories}, 
Cambridge University Press.

Cao, T.Y., ed. (1999), {\em Conceptual Foundations of Quantum Field Theory}, Cambridge 
University Press.

Cao, T.Y. (1999a), `Renormalization group: an interesting yet puzzling idea', in Cao (ed.) 
(1999), pp. 268-286.


Cao, T.Y and Schweber, S. (1993), `The conceptual foundations and the philosophical aspects 
of renormalization theory, {\em Synthese} {\bf 97}, pp. 33-108.

Castellani, E. (2002), `Reductionism, emergence and effective field theories', {\em Studies  
in the History and Philosophy of Modern Physics} {\bf 33}, 251-267.


Causey, R. (1972), `Attribute-identities and micro-reductions', {\em Journal of Philosophy} 
{\bf 67}, pp. 407-422.

Dizadji-Bahmani, F., Frigg R. and Hartmann S. (2010), `Who's afraid of Nagelian reduction?', 
{\em Erkenntnis} {\bf 73}, pp. 393-412.

Feyerabend, P. (1962), `Explanation, reduction and empiricism', in H. Feigl and G. Maxwell 
(eds.) {\em Minnesota Studies in Philosophy of Science} {\bf 3}, University of Minnesota 
Press, pp. 28-97.

Feynman, R. (1985), {\em QED}, Princeton University Press.

Fodor, J. (1974), `Special Sciences (Or: the disunity of science as a working hypothesis), 
{\em Synthese}  {\bf 28}, pp. 97-115. 

Giere, R. (1995), {\em Science Without Laws}, University of Chicago Press.

Gross, D. (1999), `Renormalization groups' in P. Deligne et al. (eds.), {\em Quantum Fields 
and Strings; a Course for Mathematicians}, American Mathematical Society; pp. 551-596.

Hartmann, S. (2001). `Effective field theories, reductionism and scientific explanation',
{\em Studies in History and Philosophy of Modern Physics} {\bf  32}, pp. 267-304.

Hempel, C. (1965), {\em Aspects of Scientific Explanation}, New York; the Free Press.

Hempel, C. (1966), {\em Philosophy of Natural Science}, Prentice-Hall.

 Hesse, M. (1965), {\em Forces and Fields}, ?Macmillan.

Horejsi, J. (1996), `Electroweak interactions and high-energy limit', {\em Czechoslovak 
Journal of Physics} {\bf 47} (1997), pp. 951-977; http://arxiv.org/abs/hep-ph/9603321

Jaffe, A. (1999), `Where does quantum field theory fit into the big picture?', in Cao (ed.) 
(1999), pp. 136-146.

Jaffe, A. (2008), `Quantum theory and relativity', in {\em Contemporary Mathematics}
(Group Representations, Ergodic Theory, and Mathematical Physics: A Tribute to George W. 
Mackey), R. Doran, C.Moore, and R. Zimmer, (eds.), {\bf 449}, pp. 209–246; available at 
http://arthurjaffe.org

Kadanoff, L. (2009), `More is the same: mean field theory and phase transitions', {\em 
Journal of Statistical Physics} {\bf 137},pp. 777-797.

Kadanoff, L. (2013), `Theories of matter: infinities and renormalization', in {\em The 
Oxford Handbook of the Philosophy of Physics}, ed. R. Batterman, Oxford University Press, 
pp. 141-188.

Kemeny, J and Oppenheim P. (1956), `On reduction', {\em Philosophical Studies} {\bf 7}, pp. 
6-19.

Kim, J. (1999), `Making sense of emergence', {\em Philosophical Studies} {\bf 95}, pp. 3-36; 
reprinted in Bedau and Humphreys (2008); page reference to the reprint.

Kim, J. (2005), {\em Physicalism, or Something Near Enough},  Princeton University Press.

Kim, J. (2006), `Emergence: Core Ideas and Issues', {\em Synthese} {\bf 151}, pp. 547-559.

Kogut, J. and Stephanov, M. (2004), {\em The Phases of Quantum Chromodynamics}, Cambridge 
University Press.

Lange, M. (2002), {\em An Introduction to the Philosophy of Physics}, Blackwell.

Lautrup, B. and Zinkernagel, H. (1999), `g-2 and the trust in experimental results', {\em 
Studies in the History and Philosophy of Modern Physics} {\bf 30}, pp. 85-110.

McMullin, E. (1993), `Indifference principle and anthropic principle in cosmology',
{\em Studies in the History and Philosophy of Science} {\bf 24}, pp. 359-389.

McMullin, E. (2002), `Origins of the field concept in physics', {\em Physics in Perspective}

Marras, A. (2002), `Kim on reduction', {\em Erkenntnis} {\bf 57}, pp. 231-257.

Menon, T. and Callender, C. (2013), `Turn and face the ch-ch-changes: philosophical 
questions raised by phase transitions', in {\em The Oxford Handbook of the Philosophy of 
Physics}, ed. R. Batterman, Oxford University Press, pp. 189-223.

Moffat, J. (2010), `Ultraviolet Complete Electroweak Model Without a Higgs Particle', {\em 
European Physics Journal Plus} {\bf 126} (2011), p. 53; http://arxiv.org/pdf/1006.1859v5

Morrison, M. (2012), `Emergent physics and micro-ontology', {\em Philosophy of Science} {\bf 
79}, pp. 141-166.

Nagel, E. (1961), {\em The Structure of Science: Problems in the Logic of Scientific 
Explanation}, Harcourt.

Nagel, E. (1979), `Issues in the logic of reductive explanations', in his {\em Teleology 
Revisited and other essays in the Philosophy and History of Science}, Columbia University 
Press; reprinted in Bedau and Humphreys (2008); page reference to the reprint.

Needham, P. (2009), `Reduction and emergence: a critique of Kim', {\em Philosophical 
Studies} {\bf 146}, pp. 93-116.

Nickles, T. (1973), `Two concepts of inter-theoretic reduction', {\em Journal of Philosophy} 
{\bf 70}, pp. 181-201.

Norton, J. (2012), `Approximation and idealization: why the difference matters', {\em 
Philosophy of Science} {\bf 74}, pp. 207-232.

Putnam, H. (1975), `Philosophy and our mental life', in his collection {\em Mind, Language 
and Reality}, Cambridge University Press, pp. 291-303.

Redhead, M. (1999), `Quantum field theory and the philosopher, in Cao (ed.) (1999), pp. 
34-40.


Schaffner, K. (1967), `Approaches to reduction', {\em Philosophy of Science} {\bf 34}, pp. 
137-147.

Schaffner, K. (1976), `Reductionism in biology: problems and prospects' in R. Cohen et al. 
(eds), {\em PSA 1974}, pp. 613-632.

Schaffner, K. (2013), `Ernest Nagel and reduction',  {\em Journal of Philosophy} {\bf 109},  
pp. 534-565.

Schweber, S. (1993), `Changing conceptualization of renormalization theory', in Brown ed. 
(1993), pp. 135-166.

Schweber, S. (1994), {\em QED and the Men who Made It}, Princeton University Press.

Sklar, L. (1967), `Types of intertheoretic reduction', {\em British Journal for the 
Philosophy of Science} {\bf 18}, pp. 109-124.

Sober, E. (1999), `The multiple realizability argument  against reductionism',  {\em 
Philosophy of Science} {\bf 66}, pp. 542-564.

Symanzik, K. (1973), `Infra-red singularities and small-distance behaviour analysis', {\em 
Communications in Mathematical Physics} {\bf 34}, pp. 7-36.

Teller, P. (1989), `Infinite renormalization', {\em Philosophy of Science} {\bf 56}, pp. 
238-257; reprinted with minor corrections as Chapter 7 of his {\em An Interpretive 
Introduction to Quantum Field Theory} (1995), Princeton University Press.


Weinberg, S. (1995), {\em The Quantum Theory of Fields}, volume 1, Cambridge University 
Press.

Weinberg, S. (1995a), {\em The Quantum Theory of Fields}, volume 2, Cambridge University 
Press.

Weinberg, S. (1999), `What is quantum field theory and what did we think it was', in Cao ed. 
(1999), pp. 241-251. Also at: arxiv: hep-th/9702027

Wightman, A. (1999), `The usefulness of a general theory of quantized fields', in Cao (ed.) 
(1999), pp. 41-46.

Wilczek, F. (2005), `Asymptotic freedom: from paradox to paradigm', (Nobel Prize Lecture 
2004), {\em Proceedings of the National Academy of Science}, {\bf 102}, pp. 8403-8413; 
available at: hep-ph/0502113

Wilson, K. (1979), `Problems in Physics with Many Scales of Length', {\em Scientific  
American} {\bf 241}, pp. 158-179.


Zuchowski (2013), `For electrodynamic consistency', forthcoming in {\em Studies in the 
History and Philosophy of Modern Physics}.

 \end{document}